\documentclass[a4paper,fleqn,usenatbib]{mnras}
\usepackage{mathptmx}
\usepackage[T1]{fontenc}
\usepackage{ae,aecompl}
\usepackage{times}
\usepackage{amssymb}
\usepackage{amsmath}
\usepackage{upgreek}
\usepackage{color}
\usepackage{graphicx}
\usepackage{rotating}
\usepackage{pdflscape}

\title[CHIMPS2]{CHIMPS2: Survey description and $^{12}$CO emission in the Galactic Centre}
\author[D.J. Eden et al.]{D. J. Eden,$^{1}$\thanks{E-mail: D.J.Eden@ljmu.ac.uk} T.J.T. Moore,$^{1}$ M.J. Currie,$^{2,3}$ A.J. Rigby,$^{4}$ E. Rosolowsky,$^{5}$ Y. Su,$^{6}$ \newauthor Kee-Tae Kim,$^{7,8}$ H. Parsons,$^{2}$ O. Morata,$^{9}$
% ...
\\
Affiliations are listed at the end of the paper}

\date{Accepted XXX. Received YYY; in original form ZZZ}

\pubyear{2020}

\begin{document}
\label{firstpage}
\pagerange{\pageref{firstpage}--\pageref{lastpage}}
\maketitle

\begin{abstract}

The latest generation of Galactic-plane surveys is enhancing our ability to study the effects of galactic environment upon the process of star formation. We present the first data from CO Heterodyne Inner Milky Way Plane Survey 2 (CHIMPS2). CHIMPS2 is a survey that will observe the Inner Galaxy, the Central Molecular Zone (CMZ), and a section of the Outer Galaxy in $^{12}$CO, $^{13}$CO, and C$^{18}$O $(J\,=\,3\rightarrow2)$ emission with the Heterodyne Array Receiver Program on the James Clerk Maxwell Telescope (JCMT). The first CHIMPS2 data presented here are a first look towards the CMZ in $^{12}$CO J\,=\,3$\rightarrow$2 and cover $-3\degr\,\leq\,\ell\,\leq\,5\degr$ and $\mid$$\emph{b}$$\mid\,\leq\,0\fdg5$ with angular resolution of 15\,arcsec, velocity resolution of 1\,km\,s$^{-1}$, and rms $\Delta\,T_A ^\ast =$ 0.58\,K at these resolutions. Such high-resolution observations of the CMZ will be a valuable data set for future studies, whilst complementing the existing Galactic Plane surveys, such as SEDIGISM, the $\emph{Herschel}$ infrared Galactic Plane Survey, and ATLASGAL. In this paper, we discuss the survey plan, the current observations and data, as well as presenting position-position maps of the region. The position-velocity maps detect foreground spiral arms in both absorption and emission.

\end{abstract}


\section{Introduction}

The formation of stars from molecular gas is the key process driving the evolution of galaxies from the early Universe to the current day. However, the regulation of the efficiency of this process (the star-formation efficiency; SFE) on both the small scales of individual clouds and the larger scales of entire galaxies, is poorly understood.

In the era of ALMA, single-dish surveys play an essential role for understanding star formation in the context of Galactic environment. Advances in array detectors have enabled large surveys of the Galactic Plane to be completed in a reasonable time, producing large samples of regions for statistical analysis \citep[e.g.,][]{Urquhart18}. By doing this, we can measure the relative impact on the SFE of Galactic-scale processes, e.g., spiral arms, or the pressure and turbulence within individual clouds.

However, untangling star formation on larger and smaller scales is complicated by the different sampling rates on these scales. Studies of extragalactic systems have produced empirical relationships, such as the Kennicutt--Schmidt (K--S) relationship \citep{Kennicutt98}, which scales the star-formation rate (SFR) with gas density; and further relationships scaling the SFR with the quantity of dense gas ($n(\rmn{H}_{\rmn{2}}) \ge 3\,\times10^{4}$\,cm$^{-3}$; \citealt{Gao04,Lada12}). These correlations, though, break down on scales of 100--500\,pc, a scale where the enclosed sample of molecular clouds is small \citep{Onodera10,Schruba10,Kruijssen14}.

These two apparently contradictory results are supported when the clump-formation efficiency (CFE), or dense-gas mass fraction (DGMF) within individual molecular clouds is examined. The distribution of cloud CFEs is lognormal, with values varying by 2--3 orders of magnitude \citep{Eden12,Eden13}; however, the CFE is fairly constant when averaged over kiloparsec scales.

The distributions of the SFEs estimated from the ratio of infrared luminosity to cloud or clump gas mass, are also found to be lognormal \citep{Eden15}, indicating that the central-limit theorem is at play in both cases, giving a well defined mean value when averaged over a large sample of clouds and a large area of the Galaxy. They also point to the spiral structures of the Milky Way having only a minor influence in enhancing the star formation within them (\citealt{Moore12}; Urquhart et al., in preparation), a conclusion also reached in M51 \citep{Schinnerer17}.  The fraction of star-forming {\em Herschel} sources as a function of Galactocentric radius in the Milky Way also displays no arm-associated signal \citep{Ragan16,Ragan18}. Studies of other Galactic-scale mechanisms, such as shear, have found conflicting evidence for impact on the star formation \citep{Dib12,Suwannajak14}.

Despite these results, there are large-scale variations between Galactic environments that would be expected to have significant influence on the star-formation process. The three major star-formation stages: the conversion of atomic to molecular gas, the conversion of molecular gas to dense star-forming clumps (DGMF and CFE), then the formation of stars (SFE), all show some significant variations related to Galactocentric radius. The molecular-gas mass fraction rapidly decreases from $\sim$\,100 per cent within the inner 1\,kpc to a few per cent at $\sim$\,10\,kpc \citep{Sofue16}. The DGMF peaks at 3--4\,kpc, and drops within the Galactic centre, where the disc may become stable against large-scale gravitational collapse \citep{Kruijssen14a}, whilst the SFE also drops dramatically in the central 0.5\,kpc when compared to the dense gas \citep{Longmore13,Urquhart13}. These reductions are within the region swept by the bar where, in external galaxies, the SFR is suppressed for the life of the bar \citep{James16,James18}. However, when compared to the total gas mass, the SFE is consistent with the K--S relationship \citep{Yusef-Zadeh09,Sormani20}.

The physics of molecular clouds are important in regulating star formation, since triggering and local environment are only thought to cause 14--30 per cent of star formation \citep{Thompson12,Kendrew12}. There is some evidence that the clouds in the Central Molecular Zone (CMZ) exhibit low SFE as they are subject to mainly solenoidal turbulence \citep{Federrath16}, as opposed to the compressive turbulence found in spiral-arm clouds. Therefore, to examine the internal physics, high-resolution observations of large samples of molecular clouds are required in different transitions and isotopologues such as the $^{13}$CO/C$^{18}$O $(J\,=\,3\rightarrow2)$ Heterodyne Inner Milky Way Plane Survey (CHIMPS; \citealt{Rigby16}), the CO High-Resolution Survey (COHRS; \citealt{Dempsey13}), the FOREST Unbiased Galactic-plane Imaging survey with the Nobeyama 45-m telescope (FUGIN; \citealt{Umemoto17}), and the Structure, Excitation, and Dynamics of the Inner Galactic Interstellar Medium survey (SEDIGISM; \citealt{Schuller17}).

CHIMPS \citep{Rigby16} was a survey covering approximately 18 square degrees of the northern inner Galactic Plane. The survey was conducted with the Heterodyne Array Receiver Program (HARP; \citealt{Buckle09}) upon the James Clerk Maxwell Telescope (JCMT) in the $J\,=\,3\rightarrow2$ rotational transitions of the CO isotopologues $^{13}$CO and C$^{18}$O, which have frequencies of 330.587\,GHz and 329.331\,GHz, respectively. The CHIMPS survey covered longitudes of $\ell$\,=\,28$\degr$--46$\degr$ at latitudes of  $\mid$\emph{b}$\mid$\,$<$\,0$\fdg$50.

COHRS \citep{Dempsey13} was also a JCMT-HARP survey of the inner Galactic Plane but in the $J\,=\,3\rightarrow2$ rotational transition of $^{12}$CO at a frequency of 345.786\,GHz. The longitude range of the initial release covers $\ell$\,=\,10$\fdg$25--55$\fdg$25, with varying latitudes between $\mid$\emph{b}$\mid$\,$<$\,0$\fdg$50 and $\mid$\emph{b}$\mid$\,$<$\,0$\fdg$25. Full coverage details and a survey description can be found in \citet{Dempsey13}.

FUGIN \citep{Umemoto17} observed the inner Galaxy ($\ell$\,=\,10$\degr$--50$\degr$, $\mid$\emph{b}$\mid$\,$<$\,1$\fdg$0) and a portion of the Outer Galaxy ($\ell$\,=\,198$\degr$--236$\degr$, $\mid$\emph{b}$\mid$\,$<$\,1$\fdg$0) using the FOREST receiver \citep{Minamidani16} upon the Nobeyama 45-m telescope in the $J\,=\,1\rightarrow0$ transition of the three isotopologues, $^{12}$CO, $^{13}$CO, and C$^{18}$O. The FUGIN survey is at an approximate resolution of 15 arcsecs, matching the CHIMPS and COHRS surveys, allowing for column density and temperatures to be calculated from a local thermodynamic equilibrium (LTE) approximation \citep{Rigby19}.

SEDIGISM \citep{Schuller17} completes the isotopologue range of CO surveys by observing $^{13}$CO and C$^{18}$O in the $J\,=\,2\rightarrow1$ rotational transition. SEDIGISM is observed at the APEX telescope at a resolution of 30\,arcsec. The longitude range is $-60\degr \le \ell \le 18\degr$, and latitude range is $\mid$\,\emph{b}\,$\mid$\,$<$\,0$\fdg$50.

The coverage of the CHIMPS, COHRS, FUGIN, and SEDIGISM surveys are summarised in Table~\ref{tab:surveys}, along with the CHIMPS2 survey regions introduced in this paper.

\begin{table*}
\begin{center}
\caption{Summary of the observation parameters for the CHIMPS, COHRS, FUGIN, and SEDIGISM surveys, including CHIMPS2 for comparison.}
\begin{minipage}{\linewidth}
\begin{tabular}{lcccccccc}
\hline
Survey & Observed & Transition & Longitude & Latitude & Angular & Velocity & Telescope & Reference$^{a}$ \\
 & Isotopologues & & Range & Range & Resolution & Resolution & & \\
\hline
CHIMPS & $^{13}$CO/C$^{18}$O & $J=3\rightarrow2$ & 28$\degr$--46$\degr$ & $\mid$\emph{b}$\mid$\,$<$\,0$\fdg$5 & 15$\arcsec$ & 0.5\,km\,s$^{-1}$ & JCMT & (1) \\
COHRS & $^{12}$CO & $J=3\rightarrow2$ & 10$\fdg$25--55$\fdg$25 & $\mid$\emph{b}$\mid$\,$<$\,0$\fdg$5 & 16$\arcsec$ & 1.0\,km\,s$^{-1}$ & JCMT & (2) \\
FUGIN Inner Gal. & $^{12}$CO/$^{13}$CO/C$^{18}$O & $J=1\rightarrow0$ & 10$\degr$--50$\degr$ & $\mid$\emph{b}$\mid$\,$<$\,1$\fdg$0 & 20$\arcsec$ & 1.3\,km\,s$^{-1}$ & NRO 45-m & (3) \\
FUGIN Outer Gal. & $^{12}$CO/$^{13}$CO/C$^{18}$O & $J=1\rightarrow0$ & 198$\degr$--236$\degr$ & $\mid$\emph{b}$\mid$\,$<$\,1$\fdg$0 & 20$\arcsec$ & 1.3\,km\,s$^{-1}$ & NRO 45-m & (3) \\
SEDIGISM & $^{13}$CO/C$^{18}$O & $J=2\rightarrow1$ & $-$60$\degr$--18$\degr$ & $\mid$\emph{b}$\mid$\,$<$\,0$\fdg$5 & 30$\arcsec$ & 0.25\,km\,s$^{-1}$ & APEX & (4) \\
\hline
CHIMPS2 CMZ & $^{12}$CO/$^{13}$CO/C$^{18}$O & $J=3\rightarrow2$ & $-$5$\degr$--5$\degr$ & $\mid$\emph{b}$\mid$\,$<$\,0$\fdg$5 & 15$\arcsec$ & 1/0.5/0.5\,km\,s$^{-1}$ & JCMT & (5) \\
CHIMPS2 Inner Gal. & $^{13}$CO/C$^{18}$O & $J=3\rightarrow2$ & 5$\degr$--28$\degr$ & $\mid$\emph{b}$\mid$\,$<$\,0$\fdg$5 & 15$\arcsec$ & 0.5\,km\,s$^{-1}$ & JCMT & (5) \\
CHIMPS2 Outer Gal. & $^{12}$CO/$^{13}$CO/C$^{18}$O & $J=3\rightarrow2$ & 215$\degr$--225$\degr$ & -2$\degr$--0$\degr$ & 15$\arcsec$ & 1/0.5/0.5\,km\,s$^{-1}$ & JCMT & (5) \\
\hline
\end{tabular}\\
$^{a}$References for survey information: (1) \citet{Rigby16}; (2) \citet{Dempsey13}; (3) \citet{Umemoto17}; (4) \citet{Schuller17}; (5) This paper.
\end{minipage}
\label{tab:surveys}
\end{center}
\end{table*}

In this paper, we describe the CHIMPS2 survey and present the first data resulting from it, being the $^{12}$CO $J\,=\,3\rightarrow2$ emission from the CMZ. The structure of this paper is as follows: Section 2 introduces the CHIMPS2 survey, the observing strategy and science goals. Section 3 describes the data and the data reduction, whilst Section 4 introduces the intensity maps from the $^{12}$CO CMZ portion of the CHIMPS2 survey, and Section 5 provides a summary.

\section{CHIMPS2}

CHIMPS2 is the follow-up to the CHIMPS and COHRS surveys and is a Large Program on the JCMT\footnote{https://www.eaobservatory.org/jcmt/science/large-programs}. The project was awarded 404 hours across four of the five JCMT weather bands to observe parts of the Inner and Outer Galaxy and the CMZ in the $J\,=\,3\rightarrow2$ transition of $^{12}$CO, $^{13}$CO, and C$^{18}$O. Table~\ref{awarded} summarises the number of hours awarded in each band. Weather Bands 1 and 2 are required for the $^{13}$CO and C$^{18}$O observations, since these transitions sit on the shoulder of the 325-GHz atmospheric water-vapour absorption feature, while Bands 4 and 5 are utilised for the $^{12}$CO data. Observations began in June 2017 and are still ongoing.

\begin{table}
\begin{center}
\caption{The time awarded to the CHIMPS2 project within each JCMT weather band, and the corresponding sky opacity.}
\label{awarded}
\begin{tabular}{cccc} \hline
Weather & Hours & Sky Opacity & CO\\
Band & Awarded & $\uptau_{225}$ & Isotopologue\\
\hline
1 &  85.5 & $<$ 0.05   & $^{13}$CO and C$^{18}$O \\
2 & 218.4 & 0.05--0.08 & $^{13}$CO and C$^{18}$O \\
4 &  50.0 & 0.12--0.20 & $^{12}$CO \\
5 &  50.0 & $>$ 0.20   & $^{12}$CO \\
\hline
\end{tabular}
\end{center}
\end{table}

\subsection{Observing Strategy}

The CHIMPS2 survey contains three components, the Inner and Outer Galaxy and the CMZ, with slightly differing observing strategies employed in each portion. The general observing strategy is to follow that of CHIMPS for $^{13}$CO and C$^{18}$O and COHRS for the $^{12}$CO observations. Full details can be found in \citet{Rigby16} and \citet{Dempsey13}; however, a brief description is included here, for completeness.

Following the CHIMPS strategy, CHIMPS2 is constructed of a grid of individual tiles orientated along Galactic coordinates.  Tiles are $21 \times 21$\,arcmin in size spaced 20\,arcmin apart, so that a $3\times3$ set of nine tiles covers an area of $\sim$\,1 square degree.  The overlap allows for calibration adjustments between tiles and correction of edge effects. The data have native angular resolution of 15\,arcsec. The $^{13}$CO and C$^{18}$O $(J\,=\,3\rightarrow2)$ lines are observed simultaneously with a 250-MHz frequency bandwidth, giving a native velocity resolution of 0.055\,km\,s$^{-1}$.  These data are binned to 0.5\,km\,s$^{-1}$, covering the $V_{\textnormal{LSR}}$ velocity ranges of $-$50 to 150\,km\,s$^{-1}$ and $-$75 to 125\,km\,s$^{-1}$, depending on the longitude of the observations. The data have antenna-temperature sensitivities of 0.58\,K and 0.73\,K in $^{13}$CO and C$^{18}$O, corresponding to H$_{2}$ column densities of 3\,$\times$\,10$^{20}$\,cm$^{-2}$ and 4\,$\times$\,10$^{21}$\,cm$^{-2}$, assuming a typical excitation temperature of 10\,K \citep[e.g.][]{Rigby19}.

The COHRS data were observed in tiles up to 0$\fdg$5\,$\times$\,0$\fdg$5 at a spatial resolution of 13.8 arcsec and a raw spectral resolution of 0.42\,km\,s$^{-1}$ in the velocity range $-$230 to 355\,km\,s$^{-1}$. The data were binned spectrally to a resolution of 0.635\,km\,s$^{-1}$. Taken across multiple weather bands, the sensitivity at this resolution is $\sim$0.3\,K (Park et al., in preparation). Since the original paper \citep{Dempsey13}, new observations have been taken to complete a uniform latitude range of $\mid\,$\emph{b}$\,\mid$\,$<$\,0$\fdg$50, to extend the longitude coverage to $\ell$\,=\,9$\fdg$50--62$\fdg$25, and to re-observe the noisiest tiles (Park et al., in preparation).

The Inner Galaxy portion of the CHIMPS2 survey is an extension of the CHIMPS and COHRS projects into the inner 3\,kpc of the Milky Way. This will extend these surveys to longitudes of $\ell$\,=\,5$\degr$ between latitudes of $\mid$\emph{b}$\mid$\,$<$\,0$\fdg$50 from their current longitude limits of $\ell$\,=\,28$\degr$ and $\ell$\,=\,10$\degr$ for CHIMPS and COHRS, respectively. The observing strategy in this region matches that of the CHIMPS and COHRS surveys, although the $^{12}$CO tiles observed in CHIMPS2 will match the $21 \times 21$\,arcmin tiles of CHIMPS.

\begin{figure*}
\begin{tabular}{ll}
\includegraphics[width=0.49\textwidth]{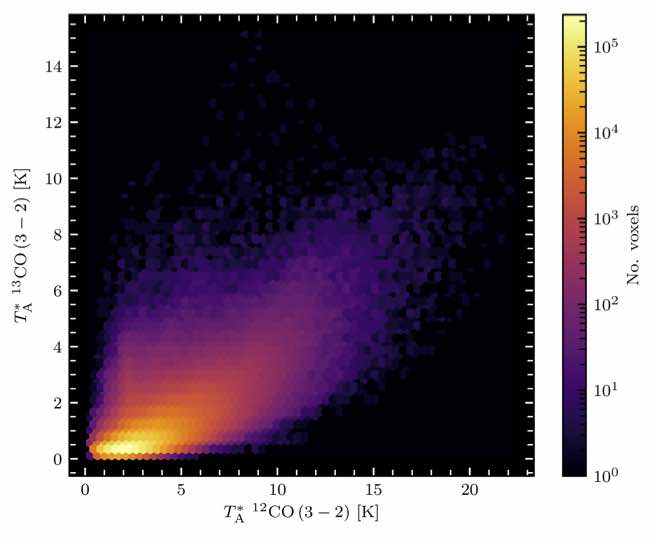} & \includegraphics[width=0.49\textwidth]{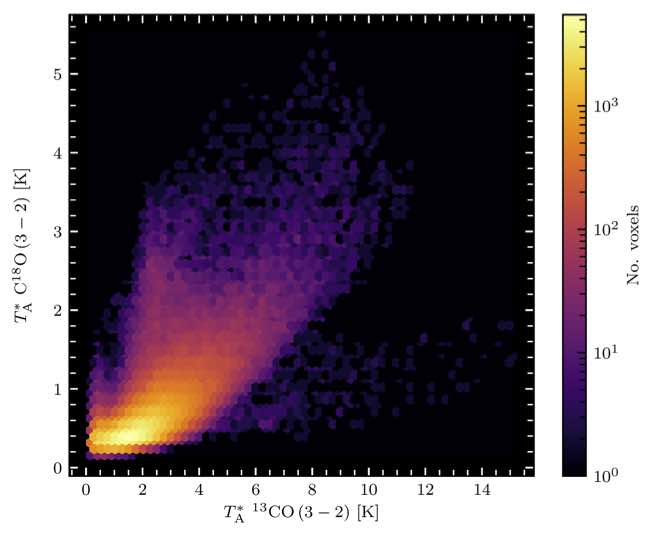} \\ 
\end{tabular}
\caption{Comparisons of brightness temperatures used to determine observing thresholds for CHIMPS2. Left panel: $^{12}$CO and $^{13}$CO $J=3 \rightarrow 2$ from COHRS and CHIMPS, respectively, used to select the detection threshold of $^{13}$CO for the Outer Galaxy segment. Right panel: $^{13}$CO and C$^{18}$O from CHIMPS used to select the detection threshold of C$^{18}$O for the CMZ segment.}
\label{thresholds}
\end{figure*}

The Outer Galaxy segment of CHIMPS2 covers the longitude and latitude ranges $215\degr \le \ell \le 225\degr$, $-2\degr \le \emph{b} \le 0\degr$, a section partly covered by the FUGIN survey and entirely by the $\emph{Herschel}$ infrared Galactic Plane Survey (Hi-GAL; \citealt{Molinari10,Molinari10a}), where over 1000 star-forming and pre-stellar clumps were identified \citep{Elia13}. This regions is also entirely covered by the Forgotten Quadrant Survey in $^{12}$CO and $^{12}$CO $J\,=\,1\rightarrow0$ \citep{Benedettini20}. The $^{12}$CO emission is, however, quite sparse in this area of the Galaxy, and a corresponding blind survey of $^{13}$CO and C$^{18}$O would result in many empty observing tiles. Therefore, using the relationship of $^{13}$CO brightness temperature from CHIMPS \citep{Rigby16} to that of $^{12}$CO from COHRS, as displayed in the left panel of Fig.~\ref{thresholds}, we are able to select regions that require $^{13}$CO and C$^{18}$O follow-up. The threshold for this was determined to be at a $^{12}$CO brightness temperature of 5\,K.

The final segment of CHIMPS2 covers the CMZ between longitudes of $\ell$\,=\,$\pm$5$\degr$ in the latitude range of $\mid\,$\emph{b}$\,\mid$\,$<$\,0$\fdg$50. This range covers the 850-$\upmu$m continuum emission presented in \citet{Parsons18}. The extended velocity range of $\sim$\,550\,km\,s$^{-1}$ present in the CO emission from the Galactic Centre \citep{Dame01}, requires the use of the 1-GHz bandwidth mode of HARP. In this mode, $^{13}$CO and C$^{18}$O cannot be observed simultaneously.  Therefore, the $^{13}$CO is observed as a blind survey, while C$^{18}$O data are taken as follow-up observations towards areas determined from the brightness-temperature relationship from CHIMPS \citep{Rigby16}, displayed in the right panel of Fig.~\ref{thresholds}. A $^{13}$CO brightness-temperature threshold of 3\,K was adopted.

\begin{figure}
\includegraphics[width=0.5\textwidth]{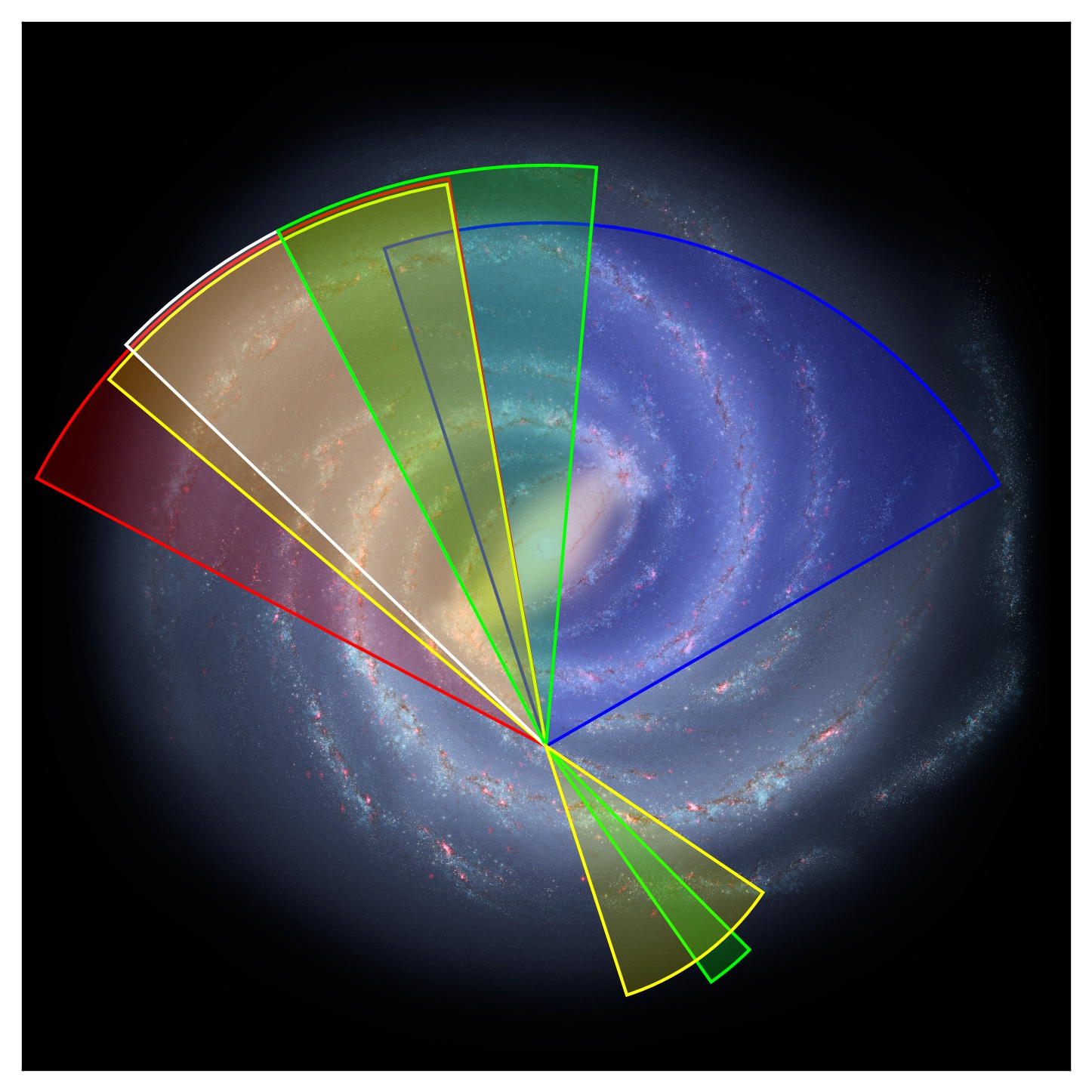}
\caption{The area of the Galaxy covered by the CHIMPS2 survey (green segments). Complementary surveys are shown for comparison of their longitude coverage, COHRS (red), CHIMPS (white), yellow (FUGIN), and SEDIGISM (blue). The background image is the artist's impression of the Milky Way by Robert Hurt of the Spitzer Science Center, made in collaboration with Robert Benjamin.}
\label{longcoverage}
\end{figure}

\begin{figure}
\includegraphics[width=0.5\textwidth]{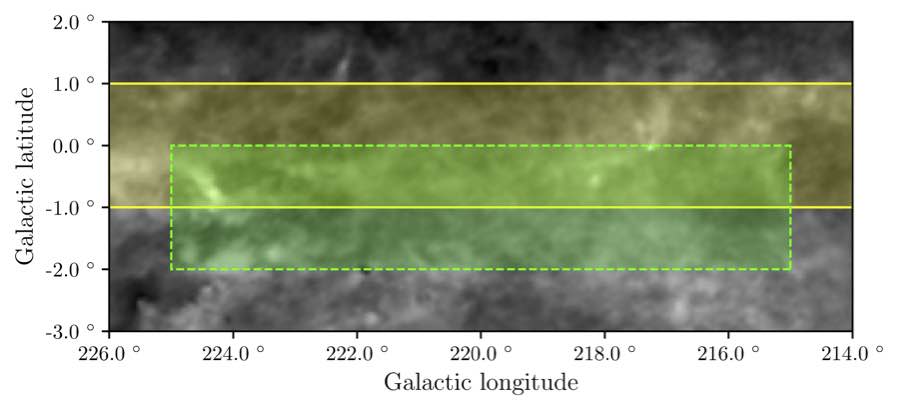}
\caption{The area of the Outer Galaxy covered by CHIMPS2 (green dashed) and FUGIN (yellow). FUGIN is extended to longitudes of $\ell$\,=\,198$\degr$ to $\ell$\,=\,236$\degr$. Hi-GAL covers the same area as CHIMPS2. The background image is the $\emph{Planck}$ dust opacity map \citep{Planck14}.}
\label{latcoverage}
\end{figure}

The longitude coverage of the CHIMPS, CHIMPS2, and COHRS surveys are shown in Fig.~\ref{longcoverage}. The FUGIN and SEDIGISM surveys are included due to the complementary nature of their observations. The CHIMPS2 latitude coverage in the Outer Galaxy follows that of Hi-GAL \citep{Molinari16} and is shown in Fig.~\ref{latcoverage}, where the FUGIN survey latitude range is also displayed. 

\subsection{Science Goals}

The science goals of the CHIMPS2 project are multi-faceted, and intended to give us a greater understanding of the effect of environment on the star-formation process. The main goals are outlined below. 
\noindent
\begin{itemize}
\item Production of comparative samples of Galactic molecular clouds across a range of Galactic environments with cloud properties, analysed using complementary CO $J\,=\,1\rightarrow0$ surveys such as FUGIN \citep{Umemoto17} and Milky Way Imaging Scroll Painting (MWISP; \citealt{Gong16,Su19}). Line-intensity ratios are found to be robust indicators of excitation conditions \citep[e.g.,][]{Nishimura15}, with simulations validating these methods \citep{Szucs14}. Multi-transition models simulating observations, such as those of \citet{Penaloza17,Penaloza18}, will refine current LTE approximate methods \citep{Rigby19}.

\item Combine with Hi-GAL \citep{Molinari16,Elia17}, JCMT Plane Survey (JPS; \citealt{Moore15,Eden17}), ATLASGAL \citep{Contreras13,Urquhart14}, and other continuum data to map the SFE and DGMF in molecular gas and constrain the mechanisms chiefly responsible for the regulation of SFE. The dense-gas SFE is largely invariant on $\sim$ kpc scales in the Inner Galaxy disc \citep{Moore12,Eden15} but falls significantly within the central 0.5\,kpc \citep{Longmore13,Urquhart13}. Comparing these regions, along with the Outer Galaxy, where the metallicity is much lower \citep{Smartt97}, and the bar-swept radii will increase our understanding of the impact of environment on the star-formation process. Variations within the CMZ may also provide insight into high-redshift star formation, since the physical condition of the clouds in this region are similar to those in galaxies at $z\,\sim\,2-3$ \citep{Kruijssen13}.

\item Analyse the turbulence within molecular clouds and its relationship to the large variations in SFE and DGMF/CFE between one cloud and another \citep{Eden12,Eden13,Eden15}. The ratio of compressive to solenoidal turbulence in molecular clouds to the CFE and SFE may determine how the internal physics of molecular clouds is altering the star formation \citep{Brunt14,Federrath16,Orkisz17}.

\item Determine Galactic structure as traced by molecular gas and star formation, and the relationship between the two. The CHIMPS survey found significant, coherent, inter-arm emission \citep{Rigby16}, identified as a connecting spur \citep{Stark06} of the type identified in external systems \citep[e.g.][]{Elmegreen80}.

\item Use comparable neutral-hydrogen data (e.g., THOR; \citealt{Beuther16}) to constrain cloud-formation models and relate turbulent conditions within molecular clouds to those in the surrounding neutral gas. The first stage of the macro star-formation process is the conversion of neutral gas into molecular gas, and therefore, clouds \citep{Wang20}. The comparison of the THOR survey with CHIMPS2 data will allow estimates of the efficiency of this process, as well as the underlying formation process \citep[e.g.][]{Bialy17} to be made.

\item Study the relationship of filaments to star formation, and of gas flow within filaments to accretion and mass accumulation in cores and clumps. The filaments in question cover different scales. Several long ($>$\,50\,pc) filamentary structures have been identified \citep{Ragan14,Zucker15}, and the CHIMPS2 data will allow for a determination of how much molecular gas is contained within these structures. On smaller scales, $\emph{Herschel}$ observations have shown a web of filamentary structures \citep[e.g.][]{Andre10,Schisano14} in which star-forming clumps are hosted \citep{Molinari10a}. The gas flow into these clumps can be traced by the high-resolution CHIMPS2 data \citep[e.g.][]{Liu18}.

\item Test current models of the gas kinematics and stability in the Galactic-centre region, the flow of gas from the disc, through the inner 3\,kpc region swept by the Galactic Bar and into the CMZ. Models of the gas flows into the centres of galaxies give signatures of these flows \citep[e.g][]{Krumholz17,Sormani19a,Armillotta19,Tress20}, and the CHIMPS2 data can determine the mass-flow rate, the nature of the flows and the star-forming properties of these clouds.

\end{itemize}

\section{Data and Data Reduction}

The data reduction for the $^{12}$CO component of the CHIMPS2 survey broadly followed the approach used for COHRS \citep{Dempsey13}, namely using the \textsc{REDUCE\_SCIENCE\_NARROWLINE} recipe of the \textsc{orac-dr} automated pipeline \citep{JenEco15}, and employing the techniques described by \cite{Jenness15}.  The pipeline invoked the Starlink applications software \citep{Currie14}, including \textsc{orac-dr}, from its 2018A release.  However, some new or improved \textsc{orac-dr} code was developed to address specific survey needs.

Since the original COHRS reductions were completed, many improvements have been made to the reduction recipe, yielding better-quality products. These include automated removal of emission from the reference (off-position) spectrum that appear as absorption lines in the reduced spectra and can bias baseline subtraction, flat fielding using a variant of the \cite{Curtis10} summation method, and masking of spectra affected by ringing in Receptor H07 \citep{Jenness15}.

The reduced spectral (position, position, velocity) cubes were re-gridded to 6-arcsec spatial pixels, convolved with a 9-arcsec Gaussian beam, resulting in  16.6-arcsec resolution. This produces an improvement on existing $^{12}$CO $(J\,=\,3\rightarrow2)$ data \citep[e.g.][]{Oka12}. Cubes with both the `native' spectral resolution and $\Delta V = 1$\,km\,s\,$^{-1}$ were generated. The cleaning came first because it included the identification and masking of spectra that contained some extraneous signal comprising alternate bright and dark spectral channels. A first-order polynomial was used to fit the baselines (aligning with COHRS; \citealt{Dempsey13}), although in the CMZ half of the baselines did require fourth-order polynomials.

The reduction of each map was made twice.  The first pass used fully automated emission detection and baseline fitting, or adopted the recipe parameters of an abutting reduced tile. A visual inspection of the resultant spectral cube, tuning through the velocities and plotting the tile's integrated spectrum, enabled refined baseline and flat-field velocity range recipe parameters to be set.  Also, any residual non-astronomical artefacts from the raw time series not removed in the quality-assurance phase of the reductions, and contamination from the off-position spectrum were assessed. In some cases of the former, such as transient narrow spikes, these were masked in the raw data before the second reduction.  Approximately 7 per cent of the tiles exhibited reference emission, which was removed by \textsc{ORAC-DR} using an algorithm that will be described in a forthcoming paper on the COHRS Second Release (Park et al., in preparation). The off-positions employed in the CHIMPS2 CMZ data are listed in Table~\ref{offpositions}.

Only 2 of 75 $^{12}$CO CMZ tiles could not be flat fielded.  In the best-determined flat fields, the corrections were typically less than 3 percent, although receptor H11 was circa 8 per cent weaker than the reference receptor. Example sets of recipe parameters are given in Appendix~\ref{recipe_param}.

All intensities given in this paper are on the $T_{\rmn{A}}^{*}$ scale. To convert this to the main-beam temperature scale, $T_{\rmn{mb}}$, use the following relation $T_{\rmn{mb}}\,=\,T_{\rmn{A}}^{*}/\upeta_{\rmn{mb}}$, where $\upeta_{\rmn{mb}}$ is the main detector efficiency and has a value of 0.72 \citep{Buckle09}.

\begin{table}
\begin{center}
\caption{The off positions for the CMZ observations in the CHIMPS2 survey.}
\label{offpositions}
\begin{tabular}{ccc} \hline
Galactic & Galactic\\
Longitude ($\degr$) & Latitude ($\degr$)\\
\hline
$-$2.50 & 2.50\\
0.78 & $-$2.75 \\
2.60 & $-$2.50\\
3.00 & $-$2.50\\
5.00 & 2.50\\
\hline
\end{tabular}
\end{center}
\end{table}

\section{Results: $^{12}$CO in the CMZ}

We are presenting the first results from the CHIMPS2 survey. These are the $^{12}$CO $J\,=\,3\rightarrow2$ emission within the CMZ. They provide a first look at the potential science that can be achieved with such data, which have greater resolution and/or trace higher densities than other large-scale CO surveys of the CMZ across the transition ladder ($J\,=\,1\rightarrow0$; \citealt{Bally87,Oka98,Dame01,Barnes15}; $J\,=\,2\rightarrow1$; \citealt{Schuller17}; $J\,=\,3\rightarrow2$; \citealt{Oka12}).
The data will be combined with the corresponding CHIMPS2 $^{13}$CO $J\,=\,3\rightarrow2$ results in a future release, along with a kinematic and dynamic analysis of the CO-traced molecular gas in the CMZ.

\begin{figure*}
\includegraphics[width=\textwidth]{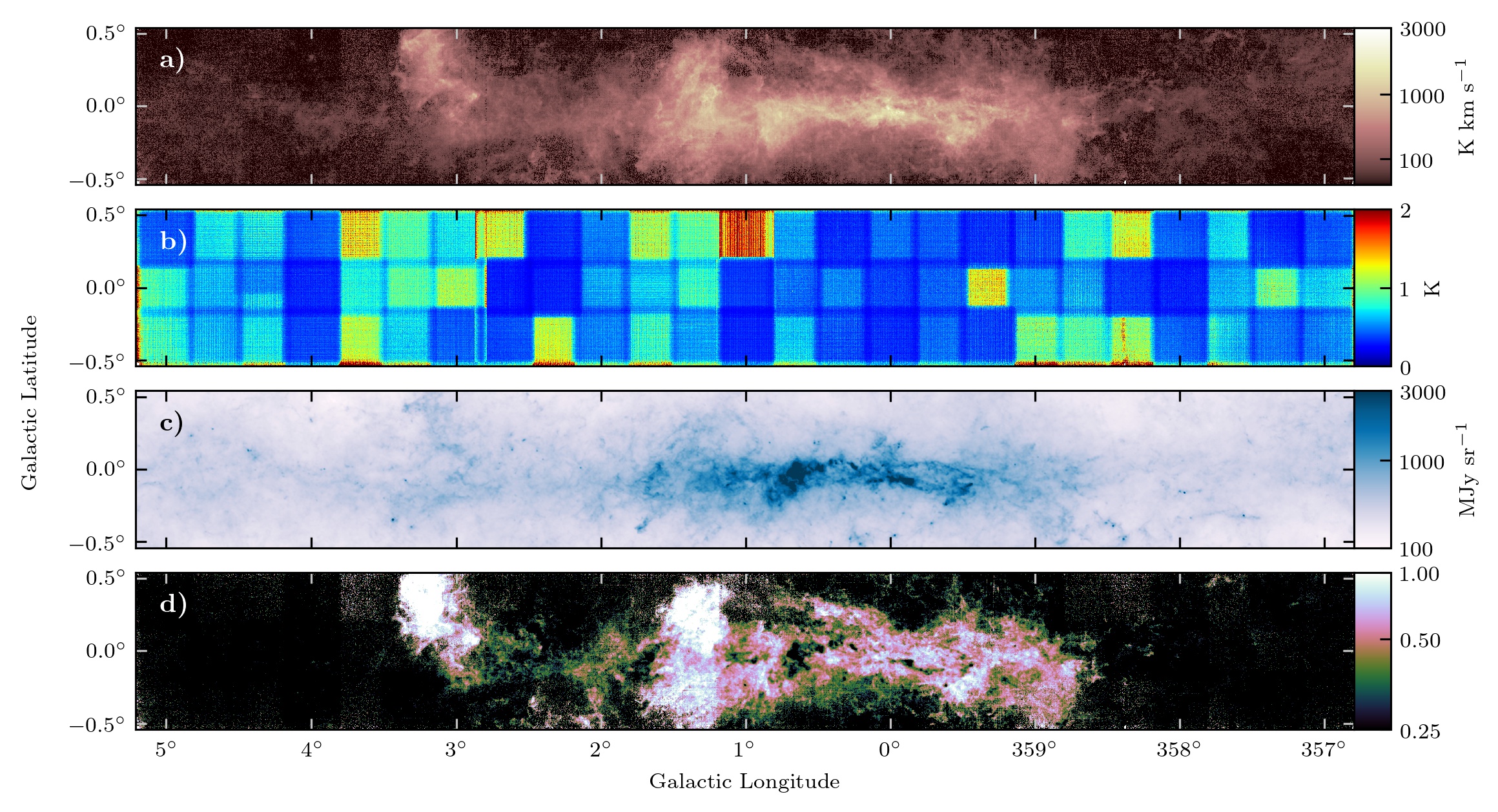}
\caption{(a) The integrated emission from the $^{12}$CO $J\,=\,3\rightarrow2$ CMZ data obtained as of October 2018. Each spectrum was integrated over all velocity channels; (b) Variance map of the $^{12}$CO $J\,=\,3\rightarrow2$ CMZ data displayed in (a); (c) \emph{Herschel} 500-$\upmu$m surface-brightness distribution from the Hi-GAL survey \citep{Molinari16}; (d) CMZ ratio of $^{12}$CO $J\,=\,3\rightarrow2$ integrated intensity (a) to \emph{Herschel} Hi-GAL 500-$\upmu$m surface brightness (c).}
\label{12COdata}
\end{figure*}

\subsection{Intensity distribution}

Panel (a) of Fig.~\ref{12COdata} shows the map of integrated intensity of $^{12}$CO $J\,=\,3\rightarrow2$ in the CMZ region between $\ell = 357^{\circ}$ and $\ell = +5^{\circ}$, $|\,b\,| \le$ 0$\fdg$5, constructed from data obtained up to the end of 2018.  
Panel (b) of Fig.~\ref{12COdata} shows the $^{12}$CO $J\,=\,3\rightarrow2$ intensity variance array mosaic and hence the relative noise levels in each constituent tile within the CMZ survey region.

\begin{figure}
\begin{tabular}{l}
\includegraphics[width=0.5\textwidth]{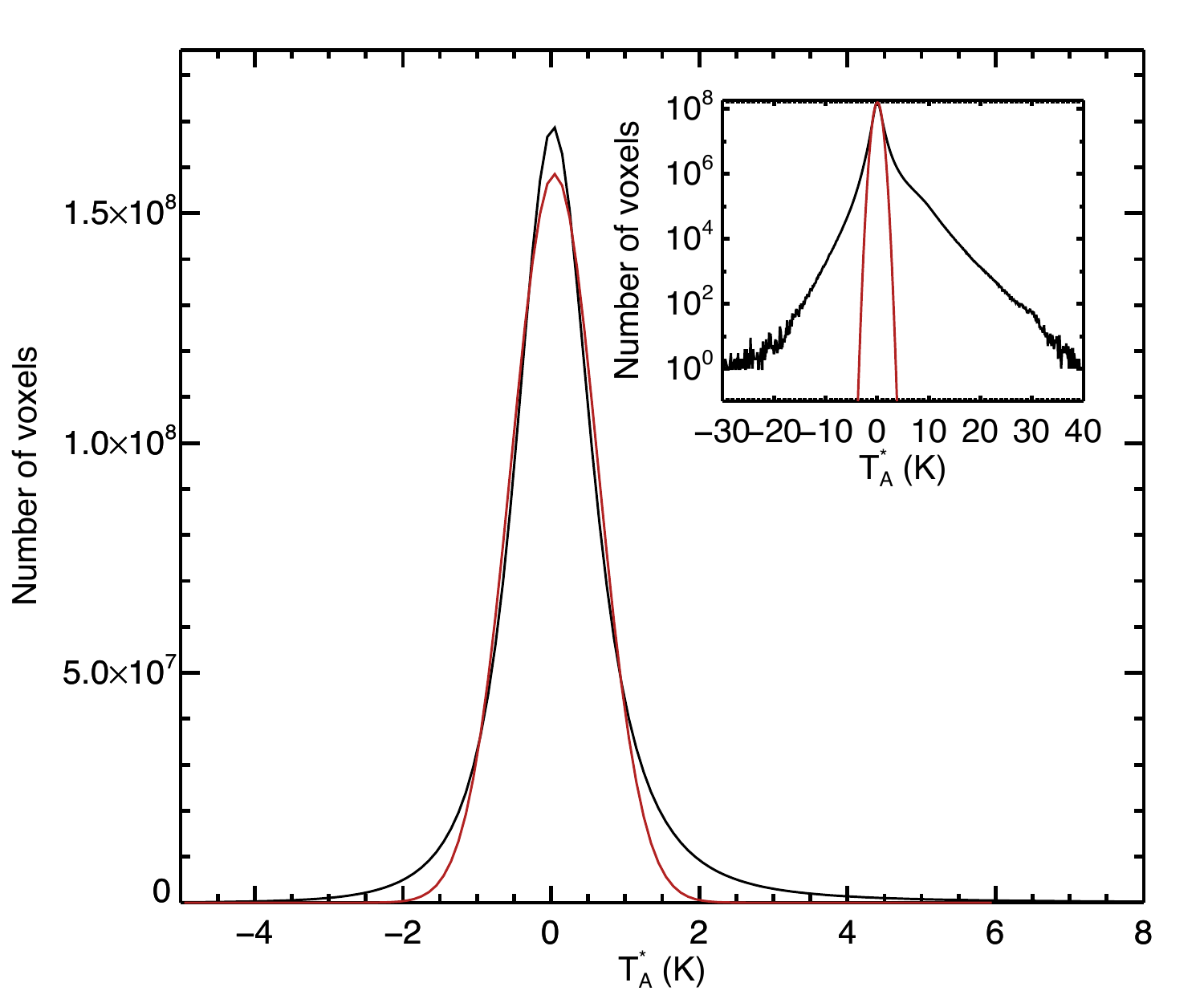} \\
\includegraphics[width=0.5\textwidth]{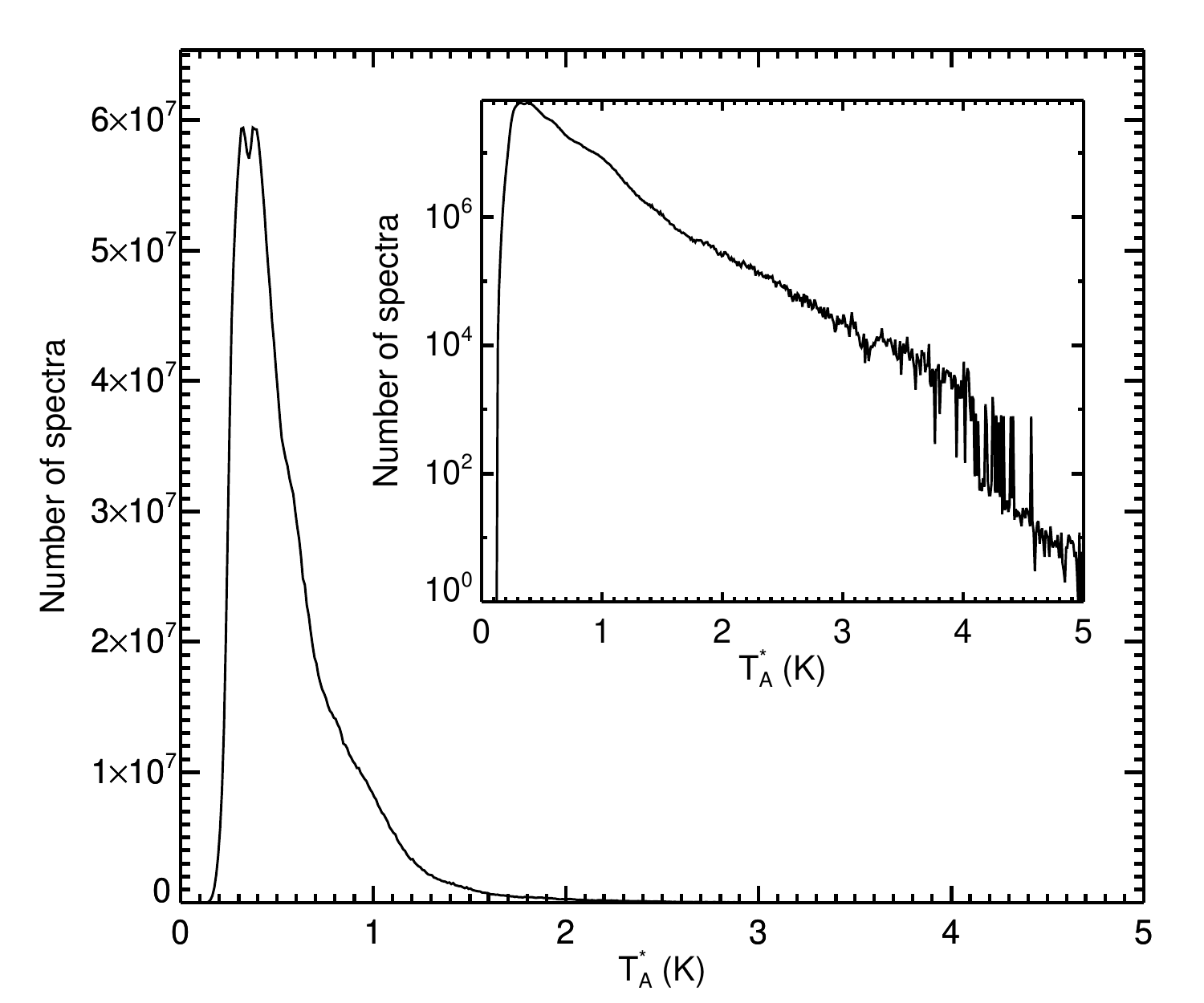}\\
\end{tabular}
\caption{Top panel: histogram of all voxels in Panel (a) of Fig.~\ref{12COdata}. The red lines display the result of a Gaussian fit to the distribution. The inset shows the distribution and Gaussian fit on a logarithmic scale. Bottom panel: histogram of the noise values in Panel (b) of Fig.~\ref{12COdata}. The double bump is due to the differing observing conditions across the map, as seen in \citet{Dempsey13}. The inset shows the same distribution on a logarithmic scale.}
\label{datahistos}
\end{figure}

A histogram of the voxel values of the map in Panel (a) of Fig.~\ref{12COdata} is displayed in the top panel of Fig.~\ref{datahistos}. The distribution is modelled by a Gaussian function with a mean of 0.05\,K and a standard deviation of 0.58\,K. The data distribution departs from the Gaussian in the negative wing due to non-Gaussian noise and non-uniform noise across the data set. In the positive wing, the excess comes from the real emission and the aforementioned noise. A histogram of the rms noise values from the variance maps in Panel (b) of Fig.~\ref{12COdata} are displayed in the bottom panel of Fig.~\ref{datahistos}. Each pixel in these variance maps represents one complete spectrum from the data cube. The values in the histogram are the square root of those in the map, giving the standard deviation. The distribution peaks at 0.38\,K, comparable with the value obtained from the Gaussian fit in the emission in the top panel of Fig.~\ref{datahistos}.

\begin{figure*}
\begin{center}
\includegraphics[width=\textwidth]{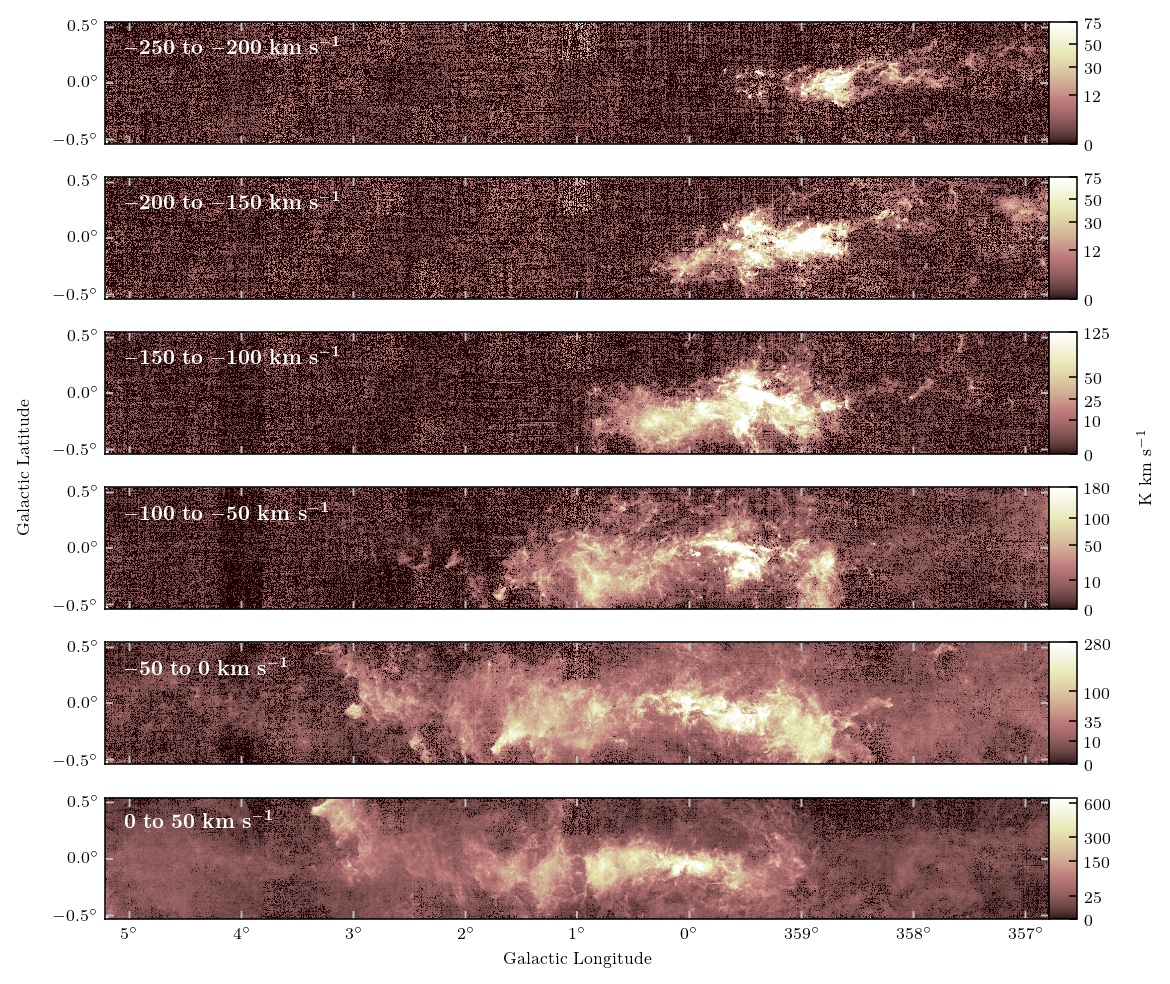}\\
\caption{The integrated emission of the map, split into 50-km\,s$^{-1}$ channels. The top map is $-$250 to $-$200\,km\,s$^{-1}$; the second map is $-$200 to $-$150\,km\,s$^{-1}$; the third map is $-$150 to $-$100\,km\,s$^{-1}$; the fourth map is $-$100 to $-$50\,km\,s$^{-1}$; the fifth map is $-$50 to 0\,km\,s$^{-1}$; and the bottom map is 0 to 50\,km\,s$^{-1}$.}
\label{channels1}
\end{center}
\end{figure*}

\begin{figure*}
\begin{center}
\includegraphics[width=\textwidth]{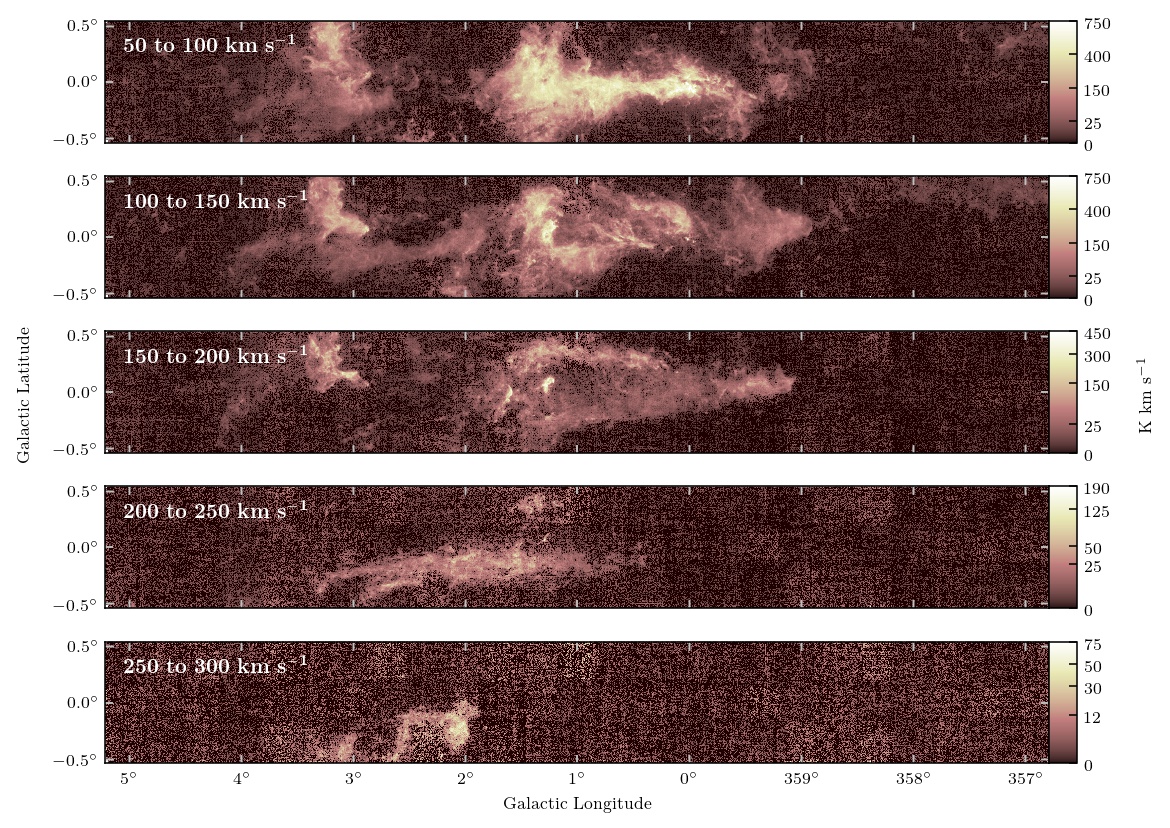}\\
\caption{The integrated emission of $^{12}$CO $J=3\rightarrow 2$, split into 50-km\,s$^{-1}$ channels.
From top to bottom, these are 50 to 100\,km\,s$^{-1}$; 100 to 150\,km\,s$^{-1}$; 150 to 200\,km\,s$^{-1}$; 200 to 250\,km\,s$^{-1}$; 250 to 300\,km\,s$^{-1}$.}
\label{channels2}
\end{center}
\end{figure*}

500-$\upmu$m continuum-emission data from the \emph{Herschel} Hi-GAL project \citep{Molinari10,Molinari11} at 37-arcsec resolution are displayed in Panel (c) of Fig.~\ref{12COdata}. Panel (d) of Fig.~\ref{12COdata} shows the distribution of the ratio of $^{12}$CO $J\,=\,3\rightarrow2$ integrated intensity to 500-$\upmu$m continuum surface brightness. The ratio values (while arbitrary) range from $\sim$ 0.1 to 2.0 -- a factor of $\sim$20.

Figs.~\ref{channels1} \& \ref{channels2} show the $^{12}$CO $J\,=\,3\rightarrow2$ emission integrated over 50-km\,s$^{-1}$ velocity windows within the range $-$250\,km\,s$^{-1}$ to 300\,km\,s$^{-1}$, with no emission detected at velocities lower than $-$250\,km\,s$^{-1}$.

\begin{figure*}
\begin{center}
\includegraphics[width=\textwidth]{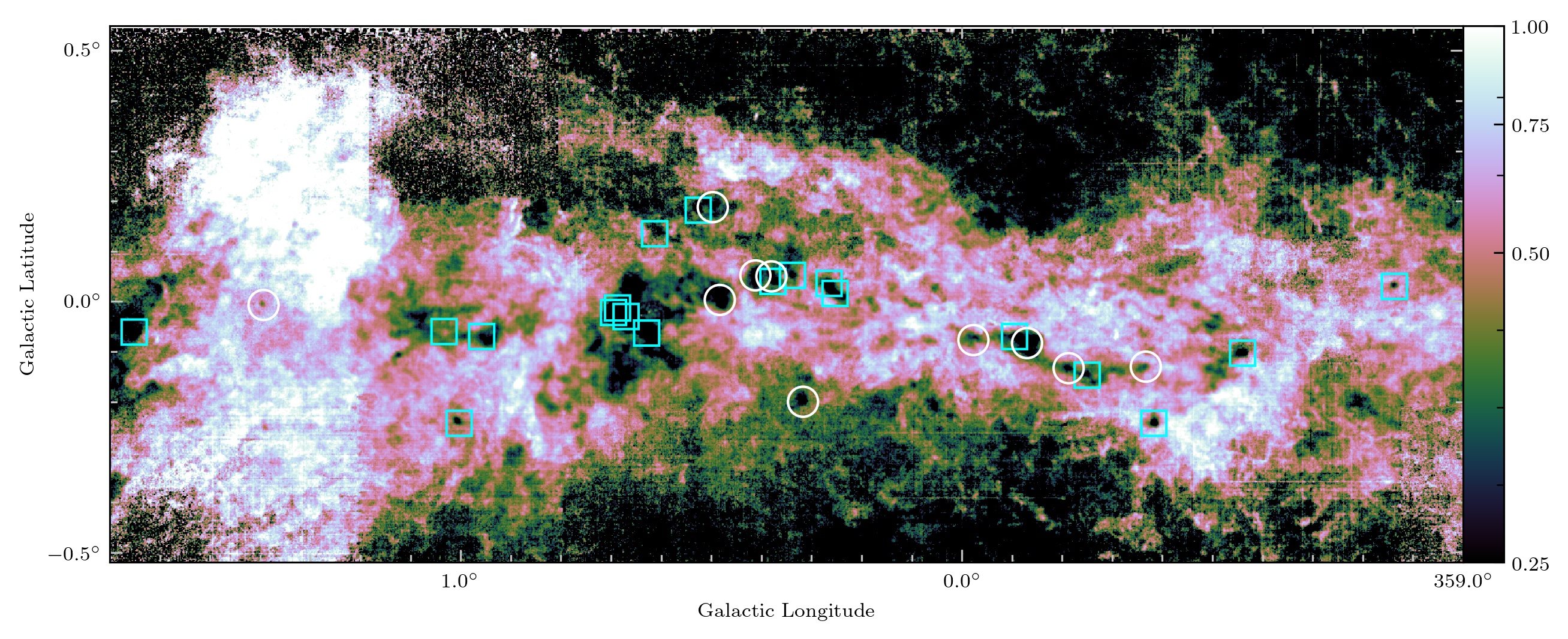}
\caption{A close-up of the central portion of Panel (d) of Fig.~\ref{12COdata}. The cyan squares are compact sources detected at 4-sigma significance using \textsc{CuTEx}.  The white circles are at the positions of several known dense clouds or clumps. Both samples are included in Table~\ref{tab:CO_500minima}.}
\label{12CO_div_500_zoom}
\end{center}
\end{figure*}

Fig.~\ref{12CO_div_500_zoom} is the same as Panel (d) in Fig.~\ref{12COdata} but with the longitude range limited to $\ell\,=\,-1\degr$ to $1\fdg7$. A number of compact minima coincident with bright regions in both the continuum and CO-line maps can be seen by eye and appear to represent high column-density objects in which the CO emission is reduced due to, e.g. high optical depth.  In order to produce an objective list of these sources, we applied the \textsc{CuTEx} object-detection package \citep{Molinari11a,Molinari17} to the inverted (reciprocal) ratio image.  \textsc{CuTEx} was chosen as it was designed to deal with extended backgrounds in \emph{Herschel} data.  The detection thresholds were four times the rms noise in the second derivative (curvature) data and a minimum of four contiguous pixels. The resulting sample was then filtered to remove sources smaller than 35\,arcsecs in either axis, to represent the 500-$\upmu$m \emph{Herschel} beam size. The detected sources are marked in Fig.~\ref{12CO_div_500_zoom} as cyan squares and listed in Table~\ref{tab:CO_500minima}. 
 
As can be seen, not all the visible compact minima were detected by \textsc{CuTEx}, including several well-known sources. Table~\ref{tab:CO_500minima} lists several of the latter that can be picked out in Fig.~\ref{12CO_div_500_zoom}, including \emph{The Brick} ($\ell \simeq 0\fdg25$), the clouds of the dust ridge at $\ell = 0\fdg3$--$0\fdg5$, Sgr B2 at $\ell \simeq 0\fdg7$, the 50- and 20-km\,s$^{-1}$ clouds at $\ell = 359\fdg9$ -- $360\fdg0$, Sgr C at $\ell \simeq 359\fdg4$, as well as the southern part of the loop structure discussed by \cite{Molinari11}, \cite{Henshaw16} and others, in terms of clouds orbiting the central potential.  The known objects from Table~\ref{tab:CO_500minima} that were not detected by \textsc{CuTEx}, are plotted in Fig.~\ref{12CO_div_500_zoom} as white circles.  In addition to these two sets of objects, there are at least as many that can be picked out by eye.  This simple analysis thus has considerable potential as a discovery channel for finding previously unknown dense, compact sources in such data and will be investigated further in future work.  Here, we briefly investigate whether or not such sources tend to be colder than their surroundings.

\begin{figure*}
\begin{center}
\begin{tabular}{ll}
\includegraphics[width=0.49\textwidth]{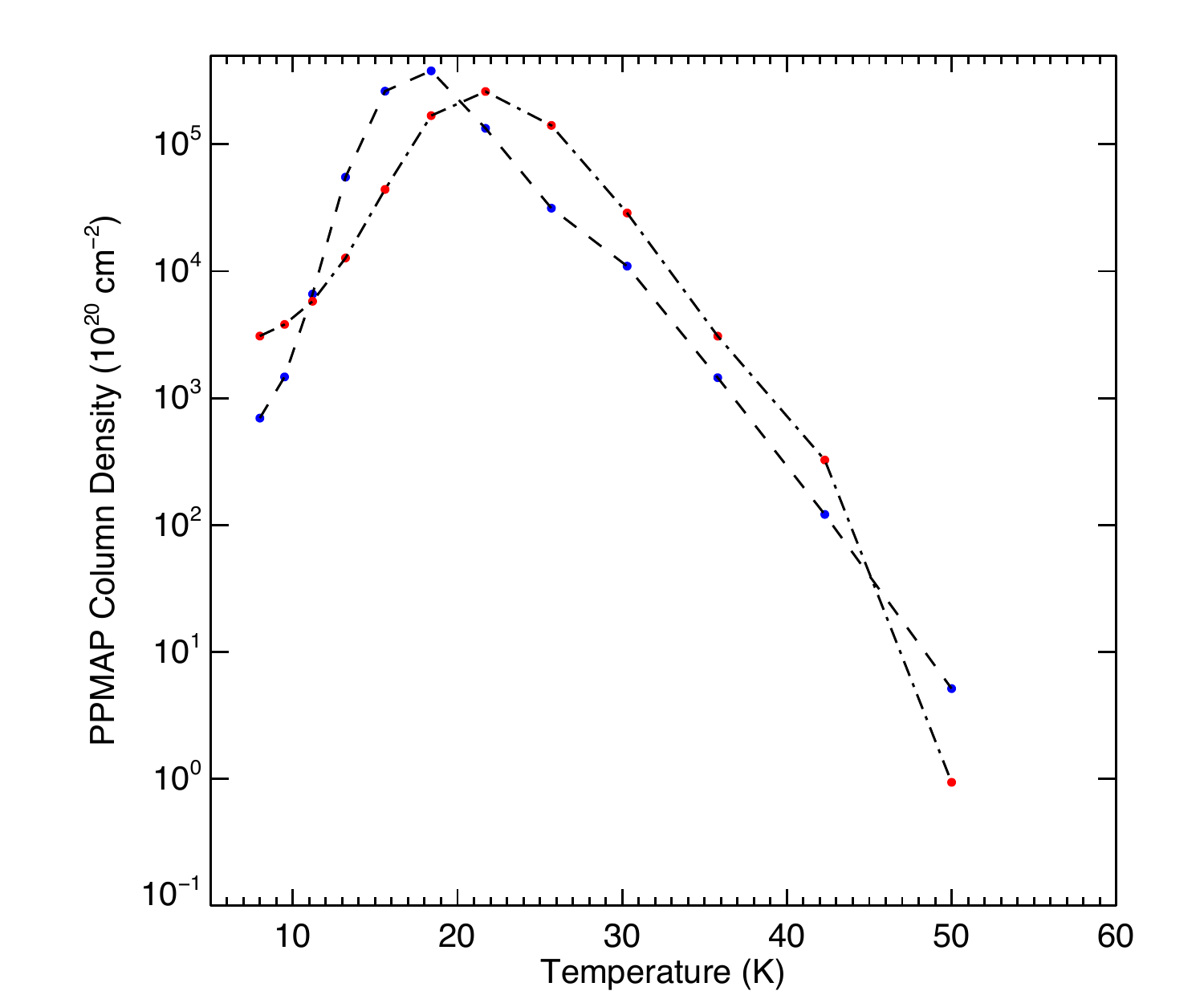} & \includegraphics[width=0.49\textwidth]{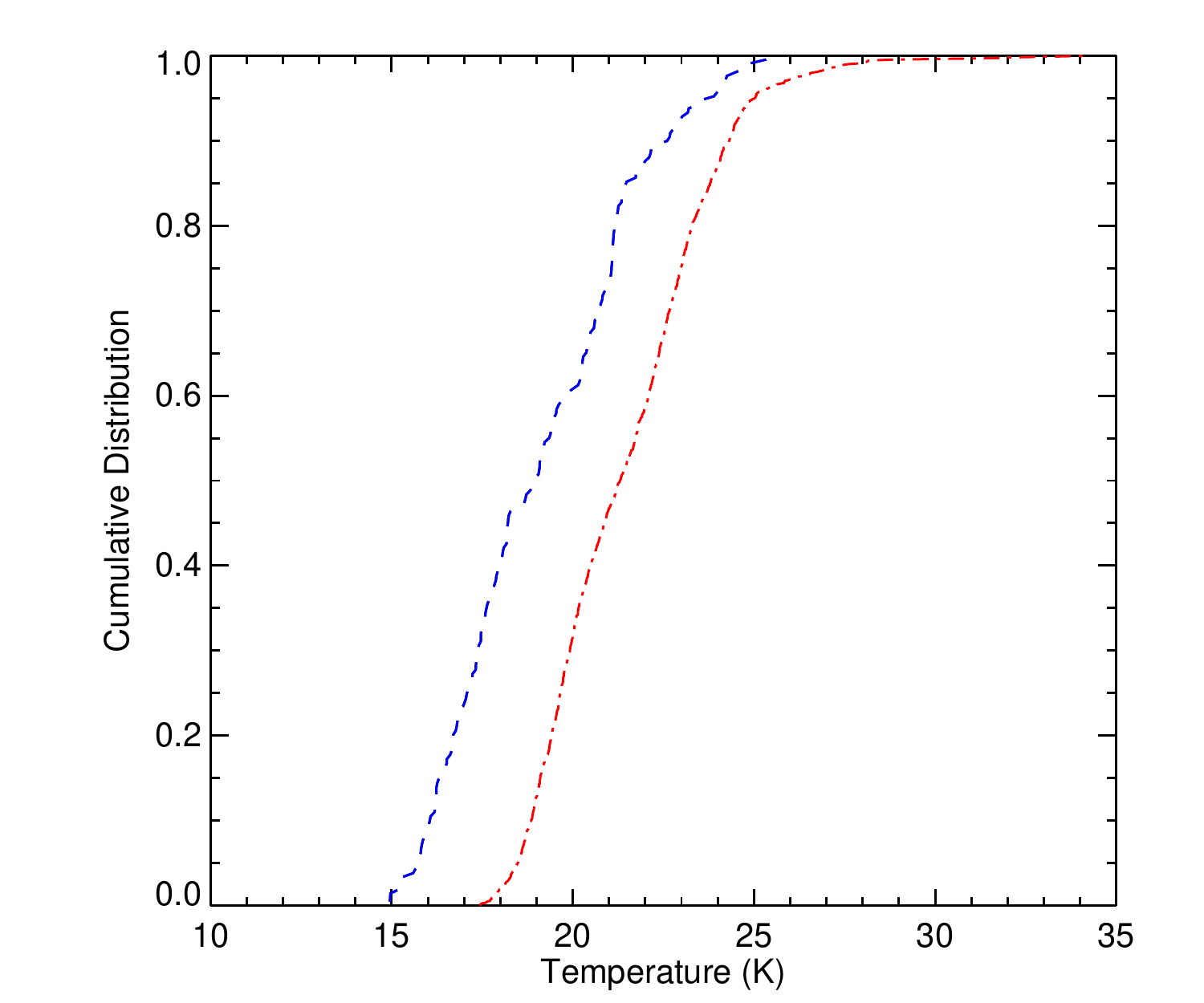} \\
\end{tabular}
\caption{Left panel: the total column density found within the \textsc{CuTEx} sources in each temperature slice from the \textsc{PPMAP} analysis of the CMZ \citep{Marsh17}. The minima from Fig.~\ref{12CO_div_500_zoom} are represented by blue points, whereas the maxima are red. Right panel: the cumulative distribution of the temperature contained within the \textsc{CuTEx} in the column-density weighted \textsc{PPMAP} temperature maps. The minima are represented by the blue dashed line, whereas the maxima are the red dot-dash line.}
\label{PPMAP}
\end{center}
\end{figure*}

The source extraction with \textsc{CuTEx} was repeated on the data in Fig.~\ref{12CO_div_500_zoom} but, rather than the reciprocal map above, now the maxima were detected. The positions of both \textsc{CuTEx} samples were used to extract temperature and column densities from the results of \citet{Marsh17}, produced by the \textsc{ppmap} procedure outlined in \citet{Marsh15}. The left panel of Fig.~\ref{PPMAP} shows the total column density contained within the sources at each temperature within the \textsc{ppmap} grid. There are 12 temperatures, evenly separated in log space between 8\,K and 50\,K. The peak total column density is found at 18.4\,K for the minima, compared with 21.7\,K for the maxima. The positions of the same sources were used to extract values from the column-density-weighted mean temperature maps produced by PPMAP, and the cumulative distributions of these values are shown in the right panel of Fig.~\ref{PPMAP}. 

The distribution of temperatures at the positions of the $^{12}$CO/500$\,\upmu$m minima in Fig.~\ref{12CO_div_500_zoom} is weighted to lower values than that of the maxima. The former are therefore tracing denser, colder structures, probably with high optical depths in $^{12}$CO and perhaps some degree of freeze-out of CO molecules onto dust grains.  The minima generally form quite compact features that pick out many of the dense clouds studied by, e.g., \cite{Walker18}.  By induction, high values, which tend to be extended, should therefore correspond to warmer areas of low $^{12}$CO optical depth.

\begin{table*}
\begin{center}
\caption{Known compact sources in the $^{12}$CO/500-$\upmu$m ratio map (Fig.~\ref{12CO_div_500_zoom}). Sources labelled with an asterisk were also detected by \textsc{CuTEx}.}
\begin{tabular}{rrll}
\hline
Galactic & Galactic & Source Name and Notes & Reference \\
Longitude ($\degr$) & Latitude ($\degr$) &  &  \\
\hline
359.137 & +0.030 & *\ion{H}{ii} region; MMB\,G359.138+00.031 & \citealt{Walsh98,Caswell10} \\
359.440 & $-$0.103 & *Sgr C & \citealt{Tsuboi91} \\
359.617 & $-$0.243 & *BGPS\,G359.617-00.243; MMB\,G359.615-00.243 & \citealt{Caswell10,Rosolowsky10} \\
359.633 & $-$0.130 & BGPS\,G359.636-00.131 & \citealt{Rosolowsky10} \\
359.750 & $-$0.147 & *AGAL\,G359.751-00.144 & \citealt{Contreras13} \\
359.787 & $-$0.133 & JCMT SCUBA source; BGPS\,G359.788-00.137 & \citealt{DiFrancesco08,Rosolowsky10} \\
359.870 & $-$0.083 & 20-km\,s$^{-1}$ cloud: UC\ion{H}{ii} regions and H$_{2}$O maser & \citealt{Downes79,Sjouwerman02} \\
359.895 & $-$0.070 & *AGAL\,G359.894$-$00.067 & \citealt{Contreras13} \\
359.977 & $-$0.077 & 50-km\,s$^{-1}$ cloud: UC\ion{H}{ii} regions and H$_{2}$O maser & \citealt{Ekers83,Reid88} \\
0.253 & +0.016 & *\emph{The Brick} & \citealt{Longmore12} \\
0.265 & +0.036 & *AGAL\,G000.264+00.032 & \citealt{Contreras13} \\
0.317 & $-$0.200 & AGAL\,0.316-0.201; MMB & \citealt{Urquhart13} \\
0.338 & +0.052 & *Dust-ridge b & \citealt{Lis99} \\
0.377 & +0.040 & *MMB\,G000.376+00.040; BGPS\,G000.378+00.041 & \citealt{Caswell10,Rosolowsky10} \\
0.380 & +0.050 & Dust-ridge c & \citealt{Lis99} \\
0.412 & +0.052 & Dust-ridge d \& BGPS\,G000.414+00.051 & \citealt{Lis99}; \citealt{Rosolowsky10} \\
0.483 & +0.003 & Sgr B1-off: UC\ion{H}{ii} regions and H$_{2}$O maser & \citealt{Lu19} \\
0.497 & +0.188 & MMB\,G000.496+00.188; BGPS\,G000.500+00.187 & \citealt{Caswell10,Rosolowsky10} \\
0.526 & +0.182 & *AGAL\,0.526+0.182 & \citealt{Contreras13} \\
0.613 & +0.135 & *2MASS\,J17463693$-$2820212 & \citealt{Cutri03} \\
0.629 & $-$0.063 & *AGAL\,G000.629$-$00.062 & \citealt{Contreras13} \\
0.670 & $-$0.030 & *Sgr B2: UC\ion{H}{ii} regions  & \citealt{Ginsburg18} \\
0.687 & $-$0.013 & *JCMT SCUBA-2 source & \citealt{Parsons18} \\
0.695 & $-$0.022 & *AGAL\,G000.693$-$00.026 & \citealt{Contreras13} \\
0.958 & $-$0.070 & *JCMT SCUBA-2 source & \citealt{Parsons18} \\
1.003 & $-$0.243 & *Sgr D1 & \citealt{Liszt92} \\
1.123 & $-$0.110 & *Sgr D UCHII + H$_2$O & \citealt{Downes66,Mehringer98} \\
1.393 & $-$0.007 & Sgr D8 & \citealt{Eckart06} \\
1.651 & $-$0.061 & *
AGAL\,G001.647$-$00.062 & \citealt{Contreras13} \\
\hline
\end{tabular}
\label{tab:CO_500minima}
\end{center}
\end{table*}

\subsection{Kinematic structure}
\label{sec:lv}

\subsubsection{High-velocity-dispersion features}

\begin{figure*}
\begin{center}
\includegraphics[width=\textwidth]{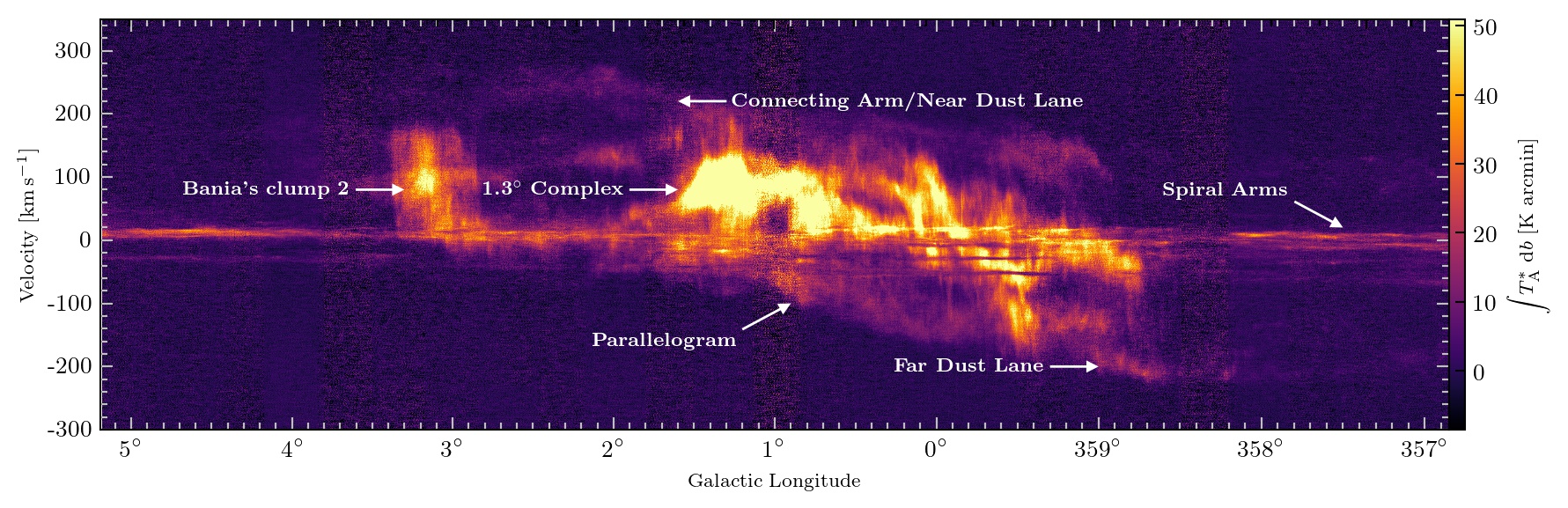}
\caption{CMZ longitude-velocity map of $^{12}$CO $J\,=\,3\rightarrow2$ intensity integrated over latitude from data complete as of Sept 2018.}
\label{12CO_CMZLV}
\end{center}
\end{figure*}

Fig.~\ref{12CO_CMZLV} contains the $\ell - V_{\textnormal{LSR}}$ distribution of the $^{12}$CO $J\,=\,3\rightarrow2$ intensity, integrated over the whole latitude range. The main features are labelled in Fig.~\ref{12CO_CMZLV} and are the parallelogram-like structure; Bania's Clump 2; the Connecting Arm, the dust lanes fuelling the CMZ; and a series of supernova remnants.

The bright, high-velocity-dispersion emission between $\ell \simeq 358\fdg5$ and $1\fdg5$; $V_{\textnormal{LSR}} \sim \pm\,250$\,km\,s$^{-1}$ in Fig.~\ref{12CO_CMZLV} that resembles a parallelogram \citep{Bania77,Bally87,Morris96} is thought to be caused by the dust lanes in the CMZ. The lateral sides are interpreted as the gas that is accreting onto the CMZ from the dust lanes \citep{Sormani19a}. The top and bottom sides are caused by gas that is partly accreting onto the CMZ after travelling past the dust lanes \citep{Sormani18}. This, combined with the efficient conversion of atomic to molecular gas, causes the velocity structure that we observe in the CMZ \citep{Sormani15}.

\begin{figure*}
\begin{center}
\includegraphics[width=\textwidth]{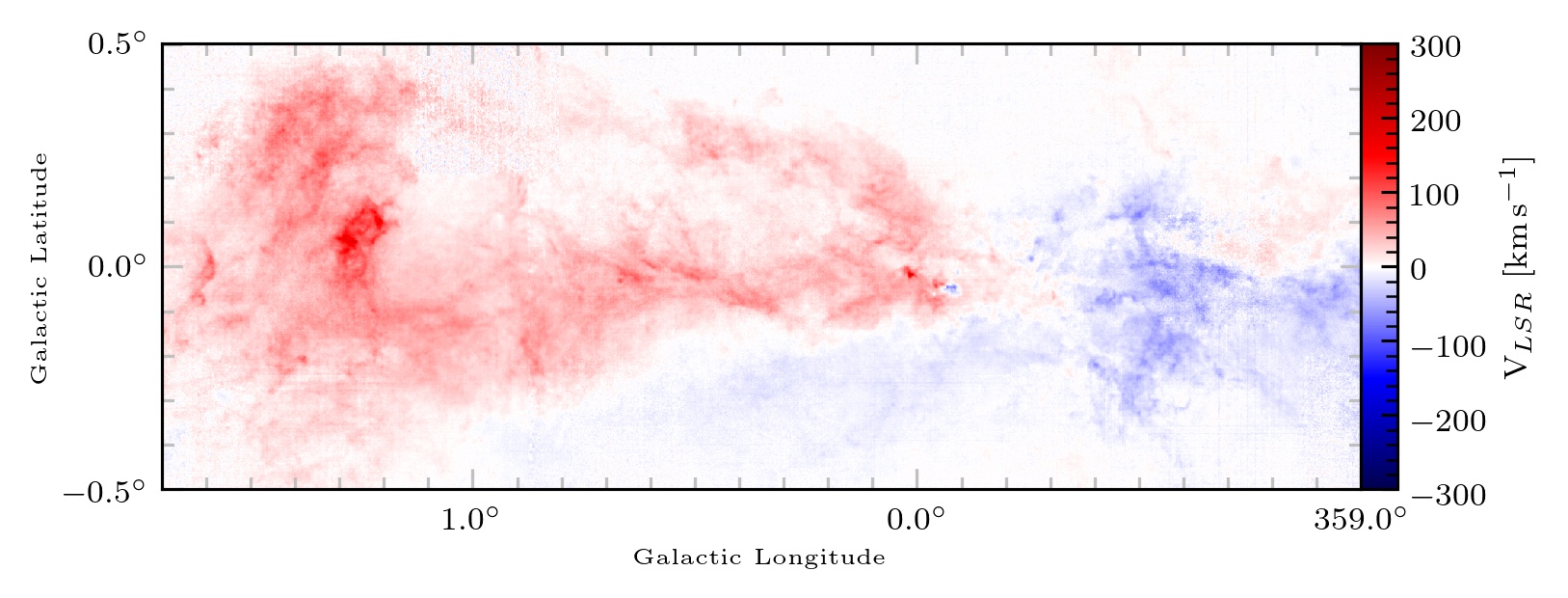}
\caption{First moment map of the $^{12}$CO CMZ map in the region represented in Fig.~\ref{12CO_div_500_zoom}.}
\label{moment_map}
\end{center}
\end{figure*}

The longitudinal asymmetry of this region of bright CO emission with respect to $\ell = 0^{\circ}$, along with the velocity centroid offset of $\sim +40$\,km\,s$^{-1}$ seen in Fig.~\ref{12CO_CMZLV}, was previously explained as the result of gas responding to an asymmetry in the Galactic potential in $m=1$ mode oscillation with respect to the Galactic disc (e.g., \citealp{Morris96}). However, the positional asymmetry has been recently suggested by \cite{Sormani18} to be due to non-steady flow of gas in the bar potential. In these models, a combination of hydrodynamical and thermal instabilities mean that the gas flow into the CMZ is clumpy and unsteady. This structure leads to transient asymmetries in the inward flow, which we observe, the authors argue, as the longitudinal asymmetry in the gas distribution. Also, structures similar to those observed at the top and bottom edges of the parallelogram feature are detected in the simulations, where they correspond to far- and near-side shocks at the leading edges of the rotating bar. The bright compact structures within this structure are the molecular clouds on librations around $x_2$ orbits in a ring around the CMZ with semi-major axis $\sim 0.3$\,kpc; and the several features that are narrow in $\ell$, but have large velocity dispersions, are shocks where the infalling material meets the CMZ or librations around an $x_2$ orbit \citep{Kruijssen15,Tress20}. The velocity offset is displayed in Fig.~\ref{moment_map}. This is the first-moment map of the sub-region in Fig.~\ref{12CO_div_500_zoom}, created using the \textsc{spectral-cube} package \citep{Ginsburg19} and reflecting the centroid velocity at each pixel.

Bania's Clump 2 can be seen as a high-velocity-dispersion cloud in Fig.~\ref{12CO_CMZLV} at $\ell$\,=\,3$\fdg$2 \citep{Bania77}. The line width of Bania's Clump 2 appears to cover over 100\,km\,s$^{-1}$ \citep{Stark86}, with very narrow longitude coverage \citep{Liszt06} but high-resolution data have found that the velocity range is made up of many lower-linewidth components \citep{Longmore17}. Clouds such as these are the signature of shocks as clouds collide with the dust lane, as opposed to the turbulence of individual clouds \citep{Sormani15c,Sormani19a}. Another high-velocity-dispersion cloud present in Fig.~\ref{12CO_CMZLV} is the $\ell$\,=\,1$\fdg$3 complex \citep{Bally88,Oka98}. The high-velocity dispersion has three potential causes. The first is a series of supernova explosions \citep{Tanaka07}, with the alternatives reflecting the acceleration of gas flows along magnetic field lines due to Parker instabilities \citep{Suzuki10,Kakiuchi18} or collisions between gas on the dust lanes and the gas orbiting the CMZ \citep{Sormani19a}. Neither of these two structures shows signatures of ongoing star formation \citep{Tanaka07,Bally10}, with no associated 70-$\upmu$m Hi-GAL compact sources \citep{Elia17}, which are considered to be a signature of active star formation \citep{Ragan16,Ragan18}.

The Connecting Arm \citep{Rodriguez-Fernandez06} is also visible in the $\ell-V_{\textnormal{LSR}}$ diagram. Though described as a spiral arm, it is in fact a dust lane at the near side of the CMZ \citep[e.g.][]{Fux99,Marshall08,Sormani18}, with a symmetrical dust lane found at the far side of the CMZ. We also see the latter in Fig.~\ref{12CO_CMZLV} as the curved feature at $V_{\textnormal{LSR}} \sim -200$\,km\,s$^{-1}$ running between $\ell \simeq 359^{\circ}$ and $357^{\circ}$. These dust lanes are signatures of accretion into the CMZ \citep{Sormani19}, fuelling episodic star formation in this region \citep{Krumholz17}.

\begin{figure}
\includegraphics[width=0.5\textwidth]{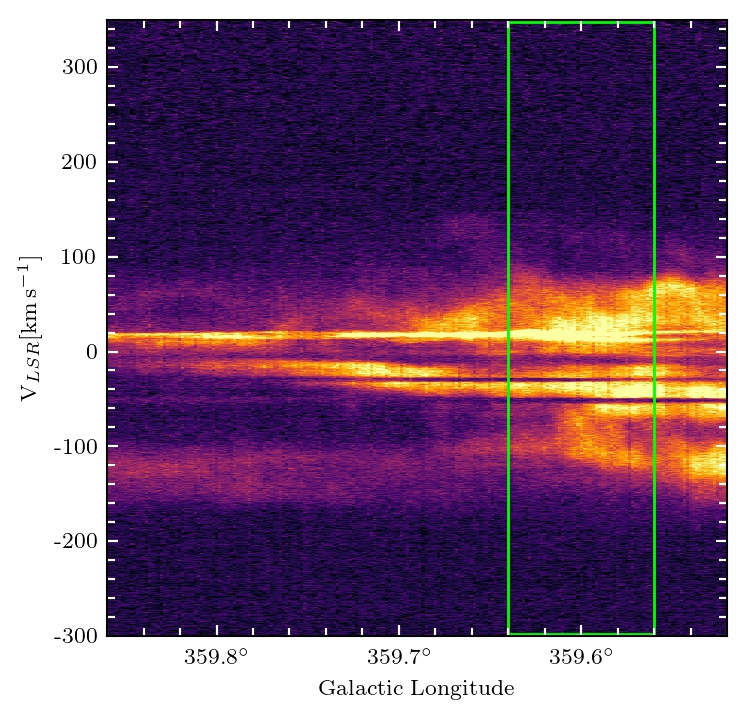}
\caption{Longitude-velocity map of the individual $^{12}$CO tile containing the reported position of the IMBH CO$-$0.40$-$0.22 \citep{Oka16,Oka17}. The expected longitude range is marked by the green rectangle.}
\label{IMBHl-v}
\end{figure}

We also confirm the findings of \citet{Tanaka18} and \citet{Reid20}, who observed no evidence of an intermediate-mass black hole (IMBH) at the position of $\ell\,=\,-0\fdg40$, $\emph{b}\,=\,-0\fdg22$ \citep{Oka16,Oka17}. Fig.~\ref{IMBHl-v} shows the $\ell - V_{\textnormal{LSR}}$ $^{12}$CO intensity distribution of the observed tile that would contain this IMBH. 
There are no large-velocity-dispersion features that are indicative of an accreting IMBH being present in the $\ell - V_{\textnormal{LSR}}$ maps.

\subsubsection{Foreground features}

\begin{figure*}
\begin{center}
\includegraphics[width=\textwidth]{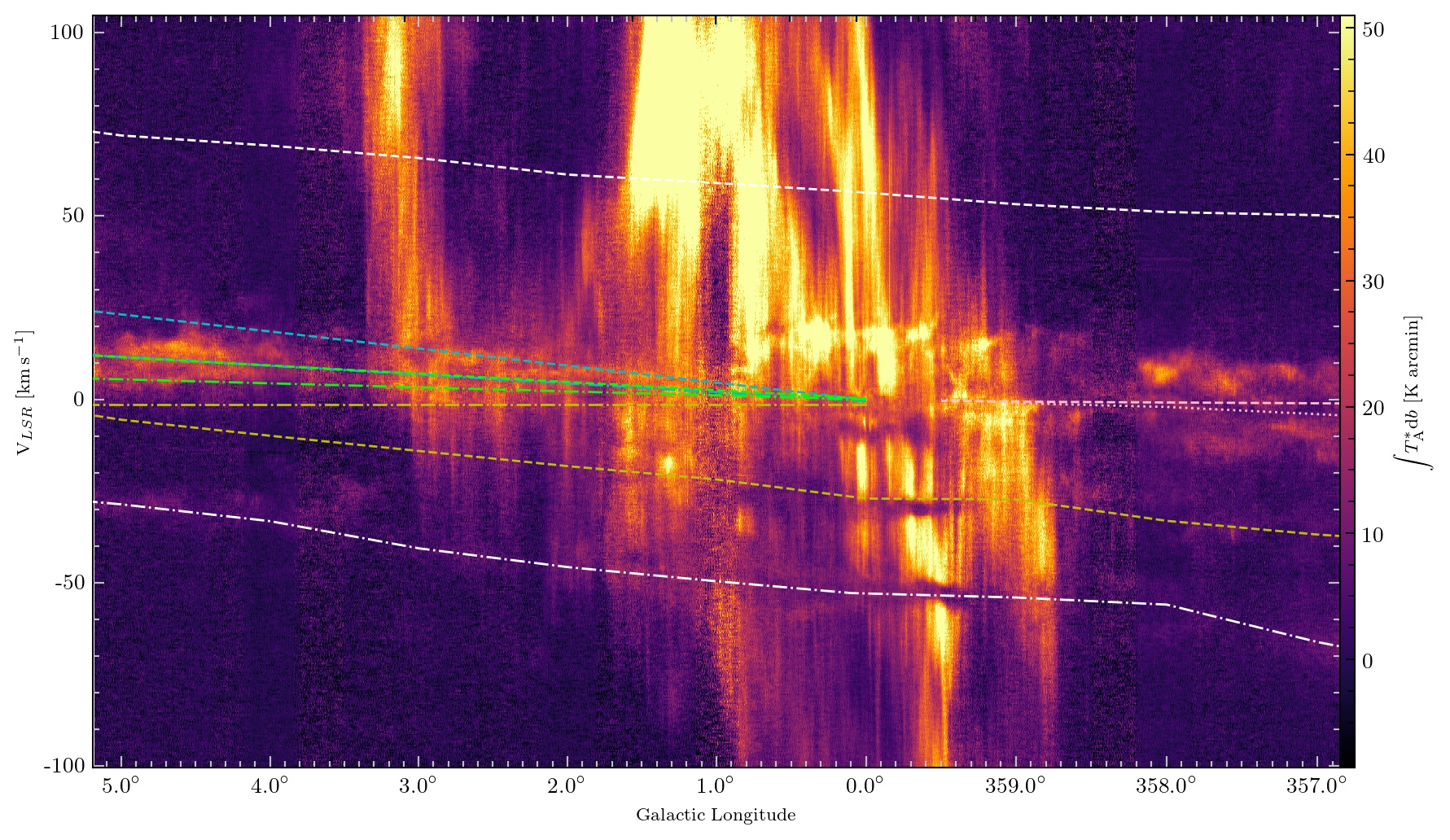}
\caption{As in Fig.~\ref{12CO_CMZLV} but with the spiral arms of \citet{Reid2016} overlaid and the velocity range restricted to $V_{\textnormal{LSR}}\,\pm\,100$\,km\,s$^{-1}$. The arm segments are labelled as follows: 3-kpc near and far arms (3kN, 3kF; white dashed and white dot-dash, respectively), Carina near portion (CrN; pink dashed), Centaurus-Crux near (CtN; pink dotted), Norma or 4-kpc (Nor; yellow dashed), Outer (Out; yellow dotted), Perseus (Per; blue dotted), Scutum near and far portions (ScN, ScF; blue dot-dash and blue dashed, respectively), and Sagittarius near and far portions (SgN, SgF; green dot-dash and green dashed, respectively). The Connecting Arm is out of this velocity range whilst the Outer Scutum-Centaurus, Carina far, an extension of the Connecting Arm, and Centaurus-Crux far arm segments currently have no parallax measurements and are not plotted.}
\label{12CO_CMZLV+arms}
\end{center}
\end{figure*}

The $\ell-V_{\textnormal{LSR}}$ plot (Fig.~\ref{12CO_CMZLV}) also shows several clear features with narrow velocity widths, in absorption and emission, probably corresponding to foreground structures, namely spiral arms. We can use these features to constrain the loci of these arms as they cross the CMZ.  Several of the arm features modelled in \cite{Reid2016} are plotted on the same data, restricted to $V_{\textnormal{LSR}}\,\pm\,100$\,km\,s$^{-1}$, in Fig.~\ref{12CO_CMZLV+arms}.

At the $\ell=0\degr$ position, there are three features in absorption at $V_{\textnormal{LSR}} \simeq -60$, $-30$ and $-10$\,km\,s$^{-1}$, with one emission feature at $\sim +10$\,km\,s$^{-1}$. All of these appear to have substructure and possibly shallow gradients and are somewhat discontinuous across the longitude range. Following \cite{Bronfman2000} and \cite{Sanna2014}, we can postulate that the $-$60\,km\,s$^{-1}$ feature is the near 3-kpc arm and the $-$30\,km\,s$^{-1}$ feature is the Norma arm.

\begin{figure*}
\begin{center}
\includegraphics[width=\textwidth]{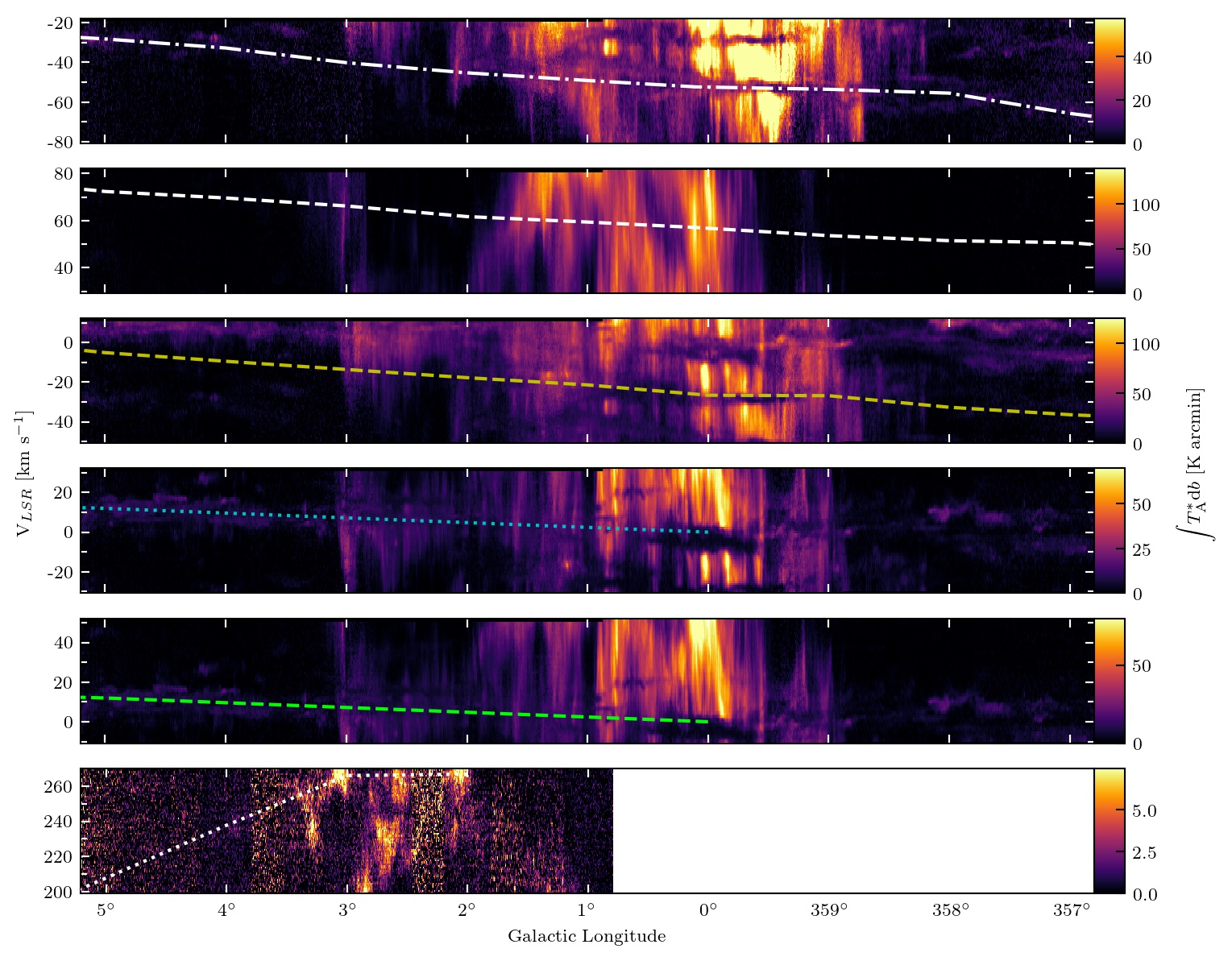}
\caption{Longitude-velocity maps isolated over the latitude and velocity range identified by \citet{Reid2016}. Top panel: near 3-kpc arm. Second panel: far 3-kpc arm. Third panel: Norma spiral arms. Fourth panel: Perseus spiral arm. Fifth panel: far Sagittarius spiral arm. Bottom panel: Connecting Arm, which is limited to a longitude range of  $\ell\,>\,0\fdg8$. The overlaid lines are the loci of the relevant spiral arms.}
\label{spiralarms}
\end{center}
\end{figure*}

To identify these features, more-precise $\ell-V_{\textnormal{LSR}}$ plots were made, integrating over the latitude and velocity range identified for these arms in \citet{Reid2016}. Fig.~\ref{spiralarms} displays the $\ell-V_{\textnormal{LSR}}$ plots for the near 3-kpc arm, far 3-kpc arm, Norma arm, Perseus arm, and the far Sagittarius arm. The latitude and velocity ranges of the five spiral arms are: $\pm$\,0$\fdg$2 and $-$80 to $-$20\,km\,s$^{-1}$, $\pm$\,0$\fdg$1 and 30 to 80\,km\,s$^{-1}$, $\pm$\,0$\fdg$2 and $-$50 to 10\,km\,s$^{-1}$, $-$0$\fdg$1 to 0$\degr$ and $-$30 to 30\,km\,s$^{-1}$, and $-$0$\fdg$1 to 0$\degr$ and $-$10 to 50\,km\,s$^{-1}$, for the near 3-kpc, far 3-kpc, Norma, Perseus, and far Sagittarius arms, respectively.

The $\ell-V_{\textnormal{LSR}}$ plots for the near 3-kpc arm and the Norma arm confirm the detection of these spiral arms. The near-3kpc arm displays absorption in the CMZ region, with emission detected in positive longitudes. The Norma spiral arm is detected in absorption. There is no evidence in these data of the far 3-kpc arm, that \cite{Sanna2014} suggest crosses $\ell = 0\degr$ at +56\,km\,s$^{-1}$.

The Perseus spiral arm and the far segment of the Sagittarius arm both have emission that corresponds to the loci of these arms, in the positive longitudes at velocities $V_{\textnormal{LSR}} \simeq +10$\,km\,s$^{-1}$. We are, therefore, unable to confirm which of these spiral arms we have detected.

We have also produced the $\ell-V_{\textnormal{LSR}}$ plot for the Connecting Arm, using the \citet{Reid2016} latitude and velocity ranges of $-$0$\fdg$5 to 0$\fdg$3 and 200 to 270\,km\,s$^{-1}$. We detect this structure, the near-side dust lane down which material streams from distances of 3\,kpc into the CMZ \citep[e.g][]{Cohen76,Rodriguez-Fernandez06,Sormani19}.

In future work, we will extract the detected narrow arm features from the $^{12}$CO data cubes in order to analyse the molecular-gas properties within them and to allow kinematic analysis of the kinematics of the residual high-velocity-dispersion emission in the CMZ itself.

\section{Summary}

We introduce the CO Heterodyne Inner Milky Way Plane Survey (CHIMPS2). CHIMPS2 will complement the CHIMPS \citep{Rigby16} and COHRS \citep{Dempsey13} surveys by observing the Central Molecular Zone (CMZ), a segment of the Outer Galaxy, and to connect the CMZ to the current CHIMPS and COHRS observations in $^{12}$CO, $^{13}$CO, and C$^{18}$O $(J\,=\,3\rightarrow2)$ emission.

We present the $^{12}$CO $J\,=\,3\rightarrow2$ data in the CMZ, covering approximately $-3\degr\,\leq\,\ell\,\leq\,5\degr$ and $\mid$$\emph{b}$$\mid\,\leq\,0\fdg50$. The data have a spatial resolution of 15\,arcsec, a spectral resolution of 1\,km\,s$^{-1}$ over velocities of $\mid$$V_{\textnormal{LSR}}$$\mid\,\leq$\,300\,km\,s$^{-1}$, an rms of 0.58\,K on 7.5\,arcsec pixels and
are available to download from the CANFAR archive. 

Taking the ratio of the integrated-intensity to the 500-$\upmu$m continuum surface brightness from Hi-GAL, we find that the result correlates well with dust temperature. The minima tend to coincide with compact, dense, cool sources; whereas the maxima correspond to warmer, more-extended regions.

We investigate the kinematic structure of the CMZ data through the use of $\ell-V_{\textnormal{LSR}}$ plots. We are able to distinguish the high-velocity-dispersion features in the Galactic Centre, such as Bania's Clump 2. We find no evidence for the existence of intermediate-mass black holes. We find evidence for spiral arms crossing in front of the Galactic Centre in both absorption and emission, detecting the near 3-kpc spiral arm, along with the Norma spiral arm, and evidence for emission in the space occupied by the far Sagittarius arm and the Perseus arm.

These data provide high-resolution observations of molecular gas in the CMZ, and will be a valuable data set for future CMZ studies, especially when combined with the future $^{13}$CO and C$^{18}$O CHIMPS2 data. Further combination with the complimentary data sets from exisiting surveys in the molecular gas, such as SEDIGISM, and in the continuum from Hi-GAL and ATLASGAL will further increase the value.

\section*{Acknowledgements}

We would like to thank the anonymous referee for their comments which have improved the clarity of the paper. DJE is supported by an STFC postdoctoral grant (ST/R000484/1). The James Clerk Maxwell Telescope is operated by the East Asian Observatory on behalf of The National Astronomical Observatory of Japan; Academia Sinica Institute of Astronomy and Astrophysics; the Korea Astronomy and Space Science Institute; the Operation, Maintenance and Upgrading Fund for Astronomical Telescopes and Facility Instruments, budgeted from the Ministry of Finance (MOF) of China and administrated by the Chinese Academy of Sciences (CAS), as well as the National Key R\&D Program of China (No. 2017YFA0402700). Additional funding support is provided by the Science and Technology Facilities Council of the United Kingdom and participating universities in the United Kingdom and Canada. The Starlink software \citep{Currie14} is currently supported by the East Asian Observatory. This research has made use of NASA's Astrophysics Data System.  MJC thanks Peter Chiu (RAL Space) for system-administrative support for the server where the data were reduced. GJW gratefully acknowledges the receipt of an Emeritus Fellowship from The Leverhulme Trust. HM and YG are supported by National Natural Science Foundation of China(NSFC) grants No. U1731237 and No. 11773054. CWL is supported by the Basic Science Research Program  through the National Research Foundation of Korea (NRF) funded by the Ministry of Education, Science and Technology (NRF-2019R1A2C1010851). 
Tie Liu is supported by international partnership program of Chinese academy of sciences grant No.114231KYSB20200009.

\section*{Data Availability}

The reduced CHIMPS2 $^{12}$CO CMZ data are available to download from the CANFAR archive\footnote{https://www.canfar.net/citation/landing?doi=20.0004}. The data are available as mosaics, roughly 2$\degr$ $\times$ 1$\degr$ in size, as well as the individual observations. Integrated $\ell - \emph{b}$ and $\ell - V_{\textnormal{LSR}}$ maps, displayed in Section 5 for the whole CMZ are provided, as well as the $\ell - V_{\textnormal{LSR}}$ maps for the individual cubes. The data are presented in FITS format.

The raw data are also downloadable from the JCMT Science Archive\footnote{http://www.cadc-ccda.hia-iha.nrc-cnrc.gc.ca/en/jcmt/} hosted by the Canadian Astronomy Data Centre using the Project ID M17BL004.

\bibliographystyle{mnras}
\bibliography{CHIMPS2_ref}

\begin{thebibliography}{}
\makeatletter
\relax
\def\mn@urlcharsother{\let\do\@makeother \do\$\do\&\do\#\do\^\do\_\do\%\do\~}
\def\mn@doi{\begingroup\mn@urlcharsother \@ifnextchar [ {\mn@doi@}
  {\mn@doi@[]}}
\def\mn@doi@[#1]#2{\def\@tempa{#1}\ifx\@tempa\@empty \href
  {http://dx.doi.org/#2} {doi:#2}\else \href {http://dx.doi.org/#2} {#1}\fi
  \endgroup}
\def\mn@eprint#1#2{\mn@eprint@#1:#2::\@nil}
\def\mn@eprint@arXiv#1{\href {http://arxiv.org/abs/#1} {{\tt arXiv:#1}}}
\def\mn@eprint@dblp#1{\href {http://dblp.uni-trier.de/rec/bibtex/#1.xml}
  {dblp:#1}}
\def\mn@eprint@#1:#2:#3:#4\@nil{\def\@tempa {#1}\def\@tempb {#2}\def\@tempc
  {#3}\ifx \@tempc \@empty \let \@tempc \@tempb \let \@tempb \@tempa \fi \ifx
  \@tempb \@empty \def\@tempb {arXiv}\fi \@ifundefined
  {mn@eprint@\@tempb}{\@tempb:\@tempc}{\expandafter \expandafter \csname
  mn@eprint@\@tempb\endcsname \expandafter{\@tempc}}}

\bibitem[\protect\citeauthoryear{{Andr{\'e}} et~al.,}{{Andr{\'e}}
  et~al.}{2010}]{Andre10}
{Andr{\'e}} P.,  et~al., 2010, \mn@doi [\aap] {10.1051/0004-6361/201014666},
  \href {http://adsabs.harvard.edu/abs/2010A%26A...518L.102A} {518, L102}

\bibitem[\protect\citeauthoryear{{Armillotta}, {Krumholz}, {Di Teodoro}  \&
  {McClure-Griffiths}}{{Armillotta} et~al.}{2019}]{Armillotta19}
{Armillotta} L.,  {Krumholz} M.~R.,  {Di Teodoro} E.~M.,   {McClure-Griffiths}
  N.~M.,  2019, \mn@doi [\mnras] {10.1093/mnras/stz2880}, \href
  {https://ui.adsabs.harvard.edu/abs/2019MNRAS.490.4401A} {490, 4401}

\bibitem[\protect\citeauthoryear{{Bally}, {Stark}, {Wilson}  \&
  {Henkel}}{{Bally} et~al.}{1987}]{Bally87}
{Bally} J.,  {Stark} A.~A.,  {Wilson} R.~W.,   {Henkel} C.,  1987, \mn@doi
  [\apjs] {10.1086/191217}, \href
  {http://adsabs.harvard.edu/abs/1987ApJS...65...13B} {65, 13}

\bibitem[\protect\citeauthoryear{{Bally}, {Stark}, {Wilson}  \&
  {Henkel}}{{Bally} et~al.}{1988}]{Bally88}
{Bally} J.,  {Stark} A.~A.,  {Wilson} R.~W.,   {Henkel} C.,  1988, \mn@doi
  [\apj] {10.1086/165891}, \href
  {http://adsabs.harvard.edu/abs/1988ApJ...324..223B} {324, 223}

\bibitem[\protect\citeauthoryear{{Bally} et~al.,}{{Bally}
  et~al.}{2010}]{Bally10}
{Bally} J.,  et~al., 2010, \mn@doi [\apj] {10.1088/0004-637X/721/1/137}, \href
  {http://adsabs.harvard.edu/abs/2010ApJ...721..137B} {721, 137}

\bibitem[\protect\citeauthoryear{{Bania}}{{Bania}}{1977}]{Bania77}
{Bania} T.~M.,  1977, \mn@doi [\apj] {10.1086/155478}, \href
  {http://adsabs.harvard.edu/abs/1977ApJ...216..381B} {216, 381}

\bibitem[\protect\citeauthoryear{{Barnes}, {Muller}, {Indermuehle},
  {O'Dougherty}, {Lowe}, {Cunningham}, {Hernandez}  \& {Fuller}}{{Barnes}
  et~al.}{2015}]{Barnes15}
{Barnes} P.~J.,  {Muller} E.,  {Indermuehle} B.,  {O'Dougherty} S.~N.,  {Lowe}
  V.,  {Cunningham} M.,  {Hernandez} A.~K.,   {Fuller} G.~A.,  2015, \mn@doi
  [\apj] {10.1088/0004-637X/812/1/6}, \href
  {https://ui.adsabs.harvard.edu/abs/2015ApJ...812....6B} {812, 6}

\bibitem[\protect\citeauthoryear{{Benedettini} et~al.,}{{Benedettini}
  et~al.}{2020}]{Benedettini20}
{Benedettini} M.,  et~al., 2020, \mn@doi [\aap] {10.1051/0004-6361/201936096},
  \href {https://ui.adsabs.harvard.edu/abs/2020A&A...633A.147B} {633, A147}

\bibitem[\protect\citeauthoryear{{Beuther} et~al.,}{{Beuther}
  et~al.}{2016}]{Beuther16}
{Beuther} H.,  et~al., 2016, \mn@doi [\aap] {10.1051/0004-6361/201629143},
  \href {http://adsabs.harvard.edu/abs/2016A%26A...595A..32B} {595, A32}

\bibitem[\protect\citeauthoryear{{Bialy}, {Bihr}, {Beuther}, {Henning}  \&
  {Sternberg}}{{Bialy} et~al.}{2017}]{Bialy17}
{Bialy} S.,  {Bihr} S.,  {Beuther} H.,  {Henning} T.,   {Sternberg} A.,  2017,
  \mn@doi [\apj] {10.3847/1538-4357/835/2/126}, \href
  {http://adsabs.harvard.edu/abs/2017ApJ...835..126B} {835, 126}

\bibitem[\protect\citeauthoryear{{Bronfman}, {Casassus}, {May}  \&
  {Nyman}}{{Bronfman} et~al.}{2000}]{Bronfman2000}
{Bronfman} L.,  {Casassus} S.,  {May} J.,   {Nyman} L.-{\AA}.,  2000, \aap,
  \href {http://adsabs.harvard.edu/abs/2000A%26A...358..521B} {358, 521}

\bibitem[\protect\citeauthoryear{{Brunt} \& {Federrath}}{{Brunt} \&
  {Federrath}}{2014}]{Brunt14}
{Brunt} C.~M.,  {Federrath} C.,  2014, \mn@doi [\mnras] {10.1093/mnras/stu888},
  \href {http://adsabs.harvard.edu/abs/2014MNRAS.442.1451B} {442, 1451}

\bibitem[\protect\citeauthoryear{{Buckle} et~al.,}{{Buckle}
  et~al.}{2009}]{Buckle09}
{Buckle} J.~V.,  et~al., 2009, \mn@doi [\mnras]
  {10.1111/j.1365-2966.2009.15347.x}, \href
  {http://adsabs.harvard.edu/abs/2009MNRAS.399.1026B} {399, 1026}

\bibitem[\protect\citeauthoryear{{Caswell} et~al.,}{{Caswell}
  et~al.}{2010}]{Caswell10}
{Caswell} J.~L.,  et~al., 2010, \mn@doi [\mnras]
  {10.1111/j.1365-2966.2010.16339.x}, \href
  {https://ui.adsabs.harvard.edu/abs/2010MNRAS.404.1029C} {404, 1029}

\bibitem[\protect\citeauthoryear{{Chapin}, {Gibb}, {Jenness}, {Berry}, {Scott}
  \& {Tilanus}}{{Chapin} et~al.}{2013}]{Chapin13}
{Chapin} E.,  {Gibb} A.~G.,  {Jenness} T.,  {Berry} D.~S.,  {Scott} D.,
  {Tilanus} R. P.~J.,  2013, Starlink User Note, \href
  {https://ui.adsabs.harvard.edu/abs/2013StaUN.258.....C} {258}

\bibitem[\protect\citeauthoryear{{Cohen} \& {Davies}}{{Cohen} \&
  {Davies}}{1976}]{Cohen76}
{Cohen} R.~J.,  {Davies} R.~D.,  1976, \mn@doi [\mnras]
  {10.1093/mnras/175.1.1}, \href
  {https://ui.adsabs.harvard.edu/abs/1976MNRAS.175....1C} {175, 1}

\bibitem[\protect\citeauthoryear{{Contreras} et~al.,}{{Contreras}
  et~al.}{2013}]{Contreras13}
{Contreras} Y.,  et~al., 2013, \mn@doi [\aap] {10.1051/0004-6361/201220155},
  \href {http://adsabs.harvard.edu/abs/2013A%26A...549A..45C} {549, A45}

\bibitem[\protect\citeauthoryear{{Currie}, {Berry}, {Jenness}, {Gibb}, {Bell}
  \& {Draper}}{{Currie} et~al.}{2014}]{Currie14}
{Currie} M.~J.,  {Berry} D.~S.,  {Jenness} T.,  {Gibb} A.~G.,  {Bell} G.~S.,
  {Draper} P.~W.,  2014, in {Manset} N.,  {Forshay} P.,  eds,  Astronomical
  Society of the Pacific Conference Series Vol. 485, Astronomical Data Analysis
  Software and Systems XXIII. p.~391

\bibitem[\protect\citeauthoryear{{Curtis}, {Richer}  \& {Buckle}}{{Curtis}
  et~al.}{2010}]{Curtis10}
{Curtis} E.~I.,  {Richer} J.~S.,   {Buckle} J.~V.,  2010, \mn@doi [MNRAS]
  {10.1111/j.1365-2966.2009.15658.x}, \href
  {http://adsabs.harvard.edu/abs/2010MNRAS.401..455C} {401, 455}

\bibitem[\protect\citeauthoryear{{Cutri} et~al.,}{{Cutri}
  et~al.}{2003}]{Cutri03}
{Cutri} R.~M.,  et~al., 2003, VizieR Online Data Catalog, \href
  {https://ui.adsabs.harvard.edu/abs/2003yCat.2246....0C} {p. II/246}

\bibitem[\protect\citeauthoryear{{Dame}, {Hartmann}  \& {Thaddeus}}{{Dame}
  et~al.}{2001}]{Dame01}
{Dame} T.~M.,  {Hartmann} D.,   {Thaddeus} P.,  2001, \mn@doi [\apj]
  {10.1086/318388}, \href {http://adsabs.harvard.edu/abs/2001ApJ...547..792D}
  {547, 792}

\bibitem[\protect\citeauthoryear{{Dempsey}, {Thomas}  \& {Currie}}{{Dempsey}
  et~al.}{2013}]{Dempsey13}
{Dempsey} J.~T.,  {Thomas} H.~S.,   {Currie} M.~J.,  2013, \mn@doi [\apjs]
  {10.1088/0067-0049/209/1/8}, \href
  {http://adsabs.harvard.edu/abs/2013ApJS..209....8D} {209, 8}

\bibitem[\protect\citeauthoryear{{Di Francesco}, {Johnstone}, {Kirk},
  {MacKenzie}  \& {Ledwosinska}}{{Di Francesco} et~al.}{2008}]{DiFrancesco08}
{Di Francesco} J.,  {Johnstone} D.,  {Kirk} H.,  {MacKenzie} T.,
  {Ledwosinska} E.,  2008, \mn@doi [\apjs] {10.1086/523645}, \href
  {https://ui.adsabs.harvard.edu/abs/2008ApJS..175..277D} {175, 277}

\bibitem[\protect\citeauthoryear{{Dib}, {Helou}, {Moore}, {Urquhart}  \&
  {Dariush}}{{Dib} et~al.}{2012}]{Dib12}
{Dib} S.,  {Helou} G.,  {Moore} T.~J.~T.,  {Urquhart} J.~S.,   {Dariush} A.,
  2012, \mn@doi [\apj] {10.1088/0004-637X/758/2/125}, \href
  {http://adsabs.harvard.edu/abs/2012ApJ...758..125D} {758, 125}

\bibitem[\protect\citeauthoryear{{Downes} \& {Maxwell}}{{Downes} \&
  {Maxwell}}{1966}]{Downes66}
{Downes} D.,  {Maxwell} A.,  1966, \mn@doi [\apj] {10.1086/148943}, \href
  {https://ui.adsabs.harvard.edu/abs/1966ApJ...146..653D} {146, 653}

\bibitem[\protect\citeauthoryear{{Downes}, {Goss}, {Schwarz}  \&
  {Wouterloot}}{{Downes} et~al.}{1979}]{Downes79}
{Downes} D.,  {Goss} W.~M.,  {Schwarz} U.~J.,   {Wouterloot} J.~G.~A.,  1979,
  \aaps, \href {https://ui.adsabs.harvard.edu/abs/1979A%26AS...35....1D} {35,
  1}

\bibitem[\protect\citeauthoryear{{Eckart} et~al.,}{{Eckart}
  et~al.}{2006}]{Eckart06}
{Eckart} A.,  et~al., 2006, \mn@doi [\aap] {10.1051/0004-6361:20054418}, \href
  {https://ui.adsabs.harvard.edu/abs/2006A%26A...450..535E} {450, 535}

\bibitem[\protect\citeauthoryear{{Eden}, {Moore}, {Plume}  \& {Morgan}}{{Eden}
  et~al.}{2012}]{Eden12}
{Eden} D.~J.,  {Moore} T.~J.~T.,  {Plume} R.,   {Morgan} L.~K.,  2012, \mn@doi
  [\mnras] {10.1111/j.1365-2966.2012.20840.x}, \href
  {http://adsabs.harvard.edu/abs/2012MNRAS.422.3178E} {422, 3178}

\bibitem[\protect\citeauthoryear{{Eden}, {Moore}, {Morgan}, {Thompson}  \&
  {Urquhart}}{{Eden} et~al.}{2013}]{Eden13}
{Eden} D.~J.,  {Moore} T.~J.~T.,  {Morgan} L.~K.,  {Thompson} M.~A.,
  {Urquhart} J.~S.,  2013, \mn@doi [\mnras] {10.1093/mnras/stt279}, \href
  {http://adsabs.harvard.edu/abs/2013MNRAS.431.1587E} {431, 1587}

\bibitem[\protect\citeauthoryear{{Eden}, {Moore}, {Urquhart}, {Elia}, {Plume},
  {Rigby}  \& {Thompson}}{{Eden} et~al.}{2015}]{Eden15}
{Eden} D.~J.,  {Moore} T.~J.~T.,  {Urquhart} J.~S.,  {Elia} D.,  {Plume} R.,
  {Rigby} A.~J.,   {Thompson} M.~A.,  2015, \mn@doi [\mnras]
  {10.1093/mnras/stv1323}, \href
  {http://adsabs.harvard.edu/abs/2015MNRAS.452..289E} {452, 289}

\bibitem[\protect\citeauthoryear{{Eden} et~al.,}{{Eden} et~al.}{2017}]{Eden17}
{Eden} D.~J.,  et~al., 2017, \mn@doi [\mnras] {10.1093/mnras/stx874}, \href
  {http://adsabs.harvard.edu/abs/2017MNRAS.469.2163E} {469, 2163}

\bibitem[\protect\citeauthoryear{{Ekers}, {van Gorkom}, {Schwarz}  \&
  {Goss}}{{Ekers} et~al.}{1983}]{Ekers83}
{Ekers} R.~D.,  {van Gorkom} J.~H.,  {Schwarz} U.~J.,   {Goss} W.~M.,  1983,
  \aap, \href {https://ui.adsabs.harvard.edu/abs/1983A&A...122..143E} {122,
  143}

\bibitem[\protect\citeauthoryear{{Elia} et~al.,}{{Elia} et~al.}{2013}]{Elia13}
{Elia} D.,  et~al., 2013, \mn@doi [\apj] {10.1088/0004-637X/772/1/45}, \href
  {http://adsabs.harvard.edu/abs/2013ApJ...772...45E} {772, 45}

\bibitem[\protect\citeauthoryear{{Elia} et~al.,}{{Elia} et~al.}{2017}]{Elia17}
{Elia} D.,  et~al., 2017, \mn@doi [\mnras] {10.1093/mnras/stx1357}, \href
  {http://adsabs.harvard.edu/abs/2017MNRAS.471..100E} {471, 100}

\bibitem[\protect\citeauthoryear{{Elmegreen}}{{Elmegreen}}{1980}]{Elmegreen80}
{Elmegreen} D.~M.,  1980, \mn@doi [\apj] {10.1086/158486}, \href
  {http://adsabs.harvard.edu/abs/1980ApJ...242..528E} {242, 528}

\bibitem[\protect\citeauthoryear{{Federrath} et~al.,}{{Federrath}
  et~al.}{2016}]{Federrath16}
{Federrath} C.,  et~al., 2016, \mn@doi [\apj] {10.3847/0004-637X/832/2/143},
  \href {http://adsabs.harvard.edu/abs/2016ApJ...832..143F} {832, 143}

\bibitem[\protect\citeauthoryear{{Fux}}{{Fux}}{1999}]{Fux99}
{Fux} R.,  1999, \aap, \href
  {http://adsabs.harvard.edu/abs/1999A%26A...345..787F} {345, 787}

\bibitem[\protect\citeauthoryear{{Gao} \& {Solomon}}{{Gao} \&
  {Solomon}}{2004}]{Gao04}
{Gao} Y.,  {Solomon} P.~M.,  2004, \mn@doi [\apj] {10.1086/382999}, \href
  {http://adsabs.harvard.edu/abs/2004ApJ...606..271G} {606, 271}

\bibitem[\protect\citeauthoryear{{Ginsburg} et~al.,}{{Ginsburg}
  et~al.}{2018}]{Ginsburg18}
{Ginsburg} A.,  et~al., 2018, \mn@doi [\apj] {10.3847/1538-4357/aaa6d4}, \href
  {https://ui.adsabs.harvard.edu/abs/2018ApJ...853..171G} {853, 171}

\bibitem[\protect\citeauthoryear{Ginsburg et~al.,}{Ginsburg
  et~al.}{2019}]{Ginsburg19}
Ginsburg A.,  et~al., 2019, radio-astro-tools/spectral-cube: v0.4.4,
  \mn@doi{10.5281/zenodo.2573901}, \url
  {https://doi.org/10.5281/zenodo.2573901}

\bibitem[\protect\citeauthoryear{{Gong} et~al.,}{{Gong} et~al.}{2016}]{Gong16}
{Gong} Y.,  et~al., 2016, \mn@doi [\aap] {10.1051/0004-6361/201527334}, \href
  {http://adsabs.harvard.edu/abs/2016A%26A...588A.104G} {588, A104}

\bibitem[\protect\citeauthoryear{{Henshaw} et~al.,}{{Henshaw}
  et~al.}{2016}]{Henshaw16}
{Henshaw} J.~D.,  et~al., 2016, \mn@doi [\mnras] {10.1093/mnras/stw121}, \href
  {https://ui.adsabs.harvard.edu/abs/2016MNRAS.457.2675H} {457, 2675}

\bibitem[\protect\citeauthoryear{{James} \& {Percival}}{{James} \&
  {Percival}}{2016}]{James16}
{James} P.~A.,  {Percival} S.~M.,  2016, \mn@doi [\mnras]
  {10.1093/mnras/stv2978}, \href
  {http://adsabs.harvard.edu/abs/2016MNRAS.457..917J} {457, 917}

\bibitem[\protect\citeauthoryear{{James} \& {Percival}}{{James} \&
  {Percival}}{2018}]{James18}
{James} P.~A.,  {Percival} S.~M.,  2018, \mn@doi [\mnras]
  {10.1093/mnras/stx2990}, \href
  {http://adsabs.harvard.edu/abs/2018MNRAS.474.3101J} {474, 3101}

\bibitem[\protect\citeauthoryear{{Jenness} \& {Economou}}{{Jenness} \&
  {Economou}}{2015}]{JenEco15}
{Jenness} T.,  {Economou} F.,  2015, \mn@doi [Astronomy and Computing]
  {10.1016/j.ascom.2014.10.005}, \href
  {https://ui.adsabs.harvard.edu/\#abs/2015A&C.....9...40J} {9, 40}

\bibitem[\protect\citeauthoryear{{Jenness}, {Chapin}, {Berry}, {Gibb},
  {Tilanus}, {Balfour}, {Tilanus}  \& {Currie}}{{Jenness}
  et~al.}{2013}]{Jenness13}
{Jenness} T.,  {Chapin} E.~L.,  {Berry} D.~S.,  {Gibb} A.~G.,  {Tilanus} R.
  P.~J.,  {Balfour} J.,  {Tilanus} V.,   {Currie} M.~J.,  2013, {SMURF:
  SubMillimeter User Reduction Facility} (\mn@eprint {ascl} {1310.007})

\bibitem[\protect\citeauthoryear{{Jenness}, {Currie}, {Tilanus}, {Cavanagh},
  {Berry}, {Leech}  \& {Rizzi}}{{Jenness} et~al.}{2015}]{Jenness15}
{Jenness} T.,  {Currie} M.~J.,  {Tilanus} R. P.~J.,  {Cavanagh} B.,  {Berry}
  D.~S.,  {Leech} J.,   {Rizzi} L.,  2015, \mn@doi [\mnras]
  {10.1093/mnras/stv1545}, \href
  {https://ui.adsabs.harvard.edu/\#abs/2015MNRAS.453...73J} {453, 73}

\bibitem[\protect\citeauthoryear{{Kakiuchi}, {Suzuki}, {Fukui}, {Torii},
  {Enokiya}, {Machida}  \& {Matsumoto}}{{Kakiuchi} et~al.}{2018}]{Kakiuchi18}
{Kakiuchi} K.,  {Suzuki} T.~K.,  {Fukui} Y.,  {Torii} K.,  {Enokiya} R.,
  {Machida} M.,   {Matsumoto} R.,  2018, \mn@doi [\mnras]
  {10.1093/mnras/sty629}, \href
  {https://ui.adsabs.harvard.edu/abs/2018MNRAS.476.5629K} {476, 5629}

\bibitem[\protect\citeauthoryear{{Kendrew} et~al.,}{{Kendrew}
  et~al.}{2012}]{Kendrew12}
{Kendrew} S.,  et~al., 2012, \mn@doi [\apj] {10.1088/0004-637X/755/1/71}, \href
  {http://adsabs.harvard.edu/abs/2012ApJ...755...71K} {755, 71}

\bibitem[\protect\citeauthoryear{{Kennicutt}}{{Kennicutt}}{1998}]{Kennicutt98}
{Kennicutt} Jr. R.~C.,  1998, \mn@doi [\apj] {10.1086/305588}, \href
  {http://adsabs.harvard.edu/abs/1998ApJ...498..541K} {498, 541}

\bibitem[\protect\citeauthoryear{{Kruijssen} \& {Longmore}}{{Kruijssen} \&
  {Longmore}}{2013}]{Kruijssen13}
{Kruijssen} J.~M.~D.,  {Longmore} S.~N.,  2013, \mn@doi [\mnras]
  {10.1093/mnras/stt1634}, \href
  {http://adsabs.harvard.edu/abs/2013MNRAS.435.2598K} {435, 2598}

\bibitem[\protect\citeauthoryear{{Kruijssen} \& {Longmore}}{{Kruijssen} \&
  {Longmore}}{2014}]{Kruijssen14}
{Kruijssen} J.~M.~D.,  {Longmore} S.~N.,  2014, \mn@doi [\mnras]
  {10.1093/mnras/stu098}, \href
  {http://adsabs.harvard.edu/abs/2014MNRAS.439.3239K} {439, 3239}

\bibitem[\protect\citeauthoryear{{Kruijssen}, {Longmore}, {Elmegreen},
  {Murray}, {Bally}, {Testi}  \& {Kennicutt}}{{Kruijssen}
  et~al.}{2014}]{Kruijssen14a}
{Kruijssen} J.~M.~D.,  {Longmore} S.~N.,  {Elmegreen} B.~G.,  {Murray} N.,
  {Bally} J.,  {Testi} L.,   {Kennicutt} R.~C.,  2014, \mn@doi [\mnras]
  {10.1093/mnras/stu494}, \href
  {http://adsabs.harvard.edu/abs/2014MNRAS.440.3370K} {440, 3370}

\bibitem[\protect\citeauthoryear{{Kruijssen}, {Dale}  \&
  {Longmore}}{{Kruijssen} et~al.}{2015}]{Kruijssen15}
{Kruijssen} J.~M.~D.,  {Dale} J.~E.,   {Longmore} S.~N.,  2015, \mn@doi
  [\mnras] {10.1093/mnras/stu2526}, \href
  {https://ui.adsabs.harvard.edu/abs/2015MNRAS.447.1059K} {447, 1059}

\bibitem[\protect\citeauthoryear{{Krumholz}, {Kruijssen}  \&
  {Crocker}}{{Krumholz} et~al.}{2017}]{Krumholz17}
{Krumholz} M.~R.,  {Kruijssen} J.~M.~D.,   {Crocker} R.~M.,  2017, \mn@doi
  [\mnras] {10.1093/mnras/stw3195}, \href
  {https://ui.adsabs.harvard.edu/abs/2017MNRAS.466.1213K} {466, 1213}

\bibitem[\protect\citeauthoryear{{Lada}, {Forbrich}, {Lombardi}  \&
  {Alves}}{{Lada} et~al.}{2012}]{Lada12}
{Lada} C.~J.,  {Forbrich} J.,  {Lombardi} M.,   {Alves} J.~F.,  2012, \mn@doi
  [\apj] {10.1088/0004-637X/745/2/190}, \href
  {http://adsabs.harvard.edu/abs/2012ApJ...745..190L} {745, 190}

\bibitem[\protect\citeauthoryear{{Lis}, {Li}, {Dowell}  \& {Menten}}{{Lis}
  et~al.}{1999}]{Lis99}
{Lis} D.~C.,  {Li} Y.,  {Dowell} C.~D.,   {Menten} K.~M.,  1999, in {Cox} P.,
  {Kessler} M.,  eds,  ESA Special Publication Vol. 427, The Universe as Seen
  by ISO. p.~627

\bibitem[\protect\citeauthoryear{{Liszt}}{{Liszt}}{1992}]{Liszt92}
{Liszt} H.~S.,  1992, \mn@doi [\apjs] {10.1086/191727}, \href
  {https://ui.adsabs.harvard.edu/abs/1992ApJS...82..495L} {82, 495}

\bibitem[\protect\citeauthoryear{{Liszt}}{{Liszt}}{2006}]{Liszt06}
{Liszt} H.~S.,  2006, \mn@doi [\aap] {10.1051/0004-6361:20054070}, \href
  {http://adsabs.harvard.edu/abs/2006A%26A...447..533L} {447, 533}

\bibitem[\protect\citeauthoryear{{Liu} et~al.,}{{Liu} et~al.}{2018}]{Liu18}
{Liu} T.,  et~al., 2018, \mn@doi [\apjs] {10.3847/1538-4365/aaa3dd}, \href
  {https://ui.adsabs.harvard.edu/abs/2018ApJS..234...28L} {234, 28}

\bibitem[\protect\citeauthoryear{{Longmore} et~al.,}{{Longmore}
  et~al.}{2012}]{Longmore12}
{Longmore} S.~N.,  et~al., 2012, \mn@doi [\apj] {10.1088/0004-637X/746/2/117},
  \href {http://adsabs.harvard.edu/abs/2012ApJ...746..117L} {746, 117}

\bibitem[\protect\citeauthoryear{{Longmore} et~al.,}{{Longmore}
  et~al.}{2013}]{Longmore13}
{Longmore} S.~N.,  et~al., 2013, \mn@doi [\mnras] {10.1093/mnras/sts376}, \href
  {https://ui.adsabs.harvard.edu/abs/2013MNRAS.429..987L} {429, 987}

\bibitem[\protect\citeauthoryear{{Longmore} et~al.,}{{Longmore}
  et~al.}{2017}]{Longmore17}
{Longmore} S.~N.,  et~al., 2017, \mn@doi [\mnras] {10.1093/mnras/stx1226},
  \href {https://ui.adsabs.harvard.edu/abs/2017MNRAS.470.1462L} {470, 1462}

\bibitem[\protect\citeauthoryear{{Lu} et~al.,}{{Lu} et~al.}{2019}]{Lu19}
{Lu} X.,  et~al., 2019, \mn@doi [\apj] {10.3847/1538-4357/ab017d}, \href
  {https://ui.adsabs.harvard.edu/abs/2019ApJ...872..171L} {872, 171}

\bibitem[\protect\citeauthoryear{{Marsh}, {Whitworth}  \& {Lomax}}{{Marsh}
  et~al.}{2015}]{Marsh15}
{Marsh} K.~A.,  {Whitworth} A.~P.,   {Lomax} O.,  2015, \mn@doi [\mnras]
  {10.1093/mnras/stv2248}, \href
  {https://ui.adsabs.harvard.edu/abs/2015MNRAS.454.4282M} {454, 4282}

\bibitem[\protect\citeauthoryear{{Marsh} et~al.,}{{Marsh}
  et~al.}{2017}]{Marsh17}
{Marsh} K.~A.,  et~al., 2017, \mn@doi [\mnras] {10.1093/mnras/stx1723}, \href
  {https://ui.adsabs.harvard.edu/abs/2017MNRAS.471.2730M} {471, 2730}

\bibitem[\protect\citeauthoryear{{Marshall}, {Fux}, {Robin}  \&
  {Reyl{\'e}}}{{Marshall} et~al.}{2008}]{Marshall08}
{Marshall} D.~J.,  {Fux} R.,  {Robin} A.~C.,   {Reyl{\'e}} C.,  2008, \mn@doi
  [\aap] {10.1051/0004-6361:20078967}, \href
  {http://adsabs.harvard.edu/abs/2008A%26A...477L..21M} {477, L21}

\bibitem[\protect\citeauthoryear{{Mehringer}, {Goss}, {Lis}, {Palmer}  \&
  {Menten}}{{Mehringer} et~al.}{1998}]{Mehringer98}
{Mehringer} D.~M.,  {Goss} W.~M.,  {Lis} D.~C.,  {Palmer} P.,   {Menten} K.~M.,
   1998, \mn@doi [\apj] {10.1086/305120}, \href
  {https://ui.adsabs.harvard.edu/abs/1998ApJ...493..274M} {493, 274}

\bibitem[\protect\citeauthoryear{{Minamidani} et~al.,}{{Minamidani}
  et~al.}{2016}]{Minamidani16}
{Minamidani} T.,  et~al., 2016, in Millimeter, Submillimeter, and Far-Infrared
  Detectors and Instrumentation for Astronomy VIII. p. 99141Z,
  \mn@doi{10.1117/12.2232137}

\bibitem[\protect\citeauthoryear{{Molinari} et~al.,}{{Molinari}
  et~al.}{2010a}]{Molinari10}
{Molinari} S.,  et~al., 2010a, \mn@doi [\pasp] {10.1086/651314}, \href
  {http://adsabs.harvard.edu/abs/2010PASP..122..314M} {122, 314}

\bibitem[\protect\citeauthoryear{{Molinari} et~al.,}{{Molinari}
  et~al.}{2010b}]{Molinari10a}
{Molinari} S.,  et~al., 2010b, \mn@doi [\aap] {10.1051/0004-6361/201014659},
  \href {http://adsabs.harvard.edu/abs/2010A%26A...518L.100M} {518, L100}

\bibitem[\protect\citeauthoryear{{Molinari}, {Schisano}, {Faustini},
  {Pestalozzi}, {di Giorgio}  \& {Liu}}{{Molinari} et~al.}{2011a}]{Molinari11a}
{Molinari} S.,  {Schisano} E.,  {Faustini} F.,  {Pestalozzi} M.,  {di Giorgio}
  A.~M.,   {Liu} S.,  2011a, \mn@doi [\aap] {10.1051/0004-6361/201014752},
  \href {https://ui.adsabs.harvard.edu/abs/2011A&A...530A.133M} {530, A133}

\bibitem[\protect\citeauthoryear{{Molinari} et~al.,}{{Molinari}
  et~al.}{2011b}]{Molinari11}
{Molinari} S.,  et~al., 2011b, \mn@doi [\apjl] {10.1088/2041-8205/735/2/L33},
  \href {http://adsabs.harvard.edu/abs/2011ApJ...735L..33M} {735, L33}

\bibitem[\protect\citeauthoryear{{Molinari} et~al.,}{{Molinari}
  et~al.}{2016}]{Molinari16}
{Molinari} S.,  et~al., 2016, \mn@doi [\aap] {10.1051/0004-6361/201526380},
  \href {http://adsabs.harvard.edu/abs/2016A%26A...591A.149M} {591, A149}

\bibitem[\protect\citeauthoryear{{Molinari}, {Schisano}, {Faustini},
  {Pestalozzi}, {di Giorgio}  \& {Liu}}{{Molinari} et~al.}{2017}]{Molinari17}
{Molinari} S.,  {Schisano} E.,  {Faustini} F.,  {Pestalozzi} M.,  {di Giorgio}
  A.~M.,   {Liu} S.,  2017, {CUTEX: CUrvature Thresholding EXtractor}
  (\mn@eprint {ascl} {1708.018})

\bibitem[\protect\citeauthoryear{{Moore}, {Urquhart}, {Morgan}  \&
  {Thompson}}{{Moore} et~al.}{2012}]{Moore12}
{Moore} T.~J.~T.,  {Urquhart} J.~S.,  {Morgan} L.~K.,   {Thompson} M.~A.,
  2012, \mn@doi [\mnras] {10.1111/j.1365-2966.2012.21740.x}, \href
  {http://adsabs.harvard.edu/abs/2012MNRAS.426..701M} {426, 701}

\bibitem[\protect\citeauthoryear{{Moore} et~al.,}{{Moore}
  et~al.}{2015}]{Moore15}
{Moore} T.~J.~T.,  et~al., 2015, \mn@doi [\mnras] {10.1093/mnras/stv1833},
  \href {http://adsabs.harvard.edu/abs/2015MNRAS.453.4264M} {453, 4264}

\bibitem[\protect\citeauthoryear{{Morris} \& {Serabyn}}{{Morris} \&
  {Serabyn}}{1996}]{Morris96}
{Morris} M.,  {Serabyn} E.,  1996, \mn@doi [\araa]
  {10.1146/annurev.astro.34.1.645}, \href
  {http://adsabs.harvard.edu/abs/1996ARA%26A..34..645M} {34, 645}

\bibitem[\protect\citeauthoryear{{Nishimura} et~al.,}{{Nishimura}
  et~al.}{2015}]{Nishimura15}
{Nishimura} A.,  et~al., 2015, \mn@doi [\apjs] {10.1088/0067-0049/216/1/18},
  \href {http://adsabs.harvard.edu/abs/2015ApJS..216...18N} {216, 18}

\bibitem[\protect\citeauthoryear{{Oka}, {Hasegawa}, {Sato}, {Tsuboi}  \&
  {Miyazaki}}{{Oka} et~al.}{1998}]{Oka98}
{Oka} T.,  {Hasegawa} T.,  {Sato} F.,  {Tsuboi} M.,   {Miyazaki} A.,  1998,
  \mn@doi [\apjs] {10.1086/313138}, \href
  {http://adsabs.harvard.edu/abs/1998ApJS..118..455O} {118, 455}

\bibitem[\protect\citeauthoryear{{Oka}, {Onodera}, {Nagai}, {Tanaka},
  {Matsumura}  \& {Kamegai}}{{Oka} et~al.}{2012}]{Oka12}
{Oka} T.,  {Onodera} Y.,  {Nagai} M.,  {Tanaka} K.,  {Matsumura} S.,
  {Kamegai} K.,  2012, \mn@doi [\apjs] {10.1088/0067-0049/201/2/14}, \href
  {https://ui.adsabs.harvard.edu/abs/2012ApJS..201...14O} {201, 14}

\bibitem[\protect\citeauthoryear{{Oka}, {Mizuno}, {Miura}  \& {Takekawa}}{{Oka}
  et~al.}{2016}]{Oka16}
{Oka} T.,  {Mizuno} R.,  {Miura} K.,   {Takekawa} S.,  2016, \mn@doi [\apjl]
  {10.3847/2041-8205/816/1/L7}, \href
  {https://ui.adsabs.harvard.edu/abs/2016ApJ...816L...7O} {816, L7}

\bibitem[\protect\citeauthoryear{{Oka}, {Tsujimoto}, {Iwata}, {Nomura}  \&
  {Takekawa}}{{Oka} et~al.}{2017}]{Oka17}
{Oka} T.,  {Tsujimoto} S.,  {Iwata} Y.,  {Nomura} M.,   {Takekawa} S.,  2017,
  \mn@doi [Nature Astronomy] {10.1038/s41550-017-0224-z}, \href
  {https://ui.adsabs.harvard.edu/abs/2017NatAs...1..709O} {1, 709}

\bibitem[\protect\citeauthoryear{{Onodera} et~al.,}{{Onodera}
  et~al.}{2010}]{Onodera10}
{Onodera} S.,  et~al., 2010, \mn@doi [\apjl] {10.1088/2041-8205/722/2/L127},
  \href {https://ui.adsabs.harvard.edu/abs/2010ApJ...722L.127O} {722, L127}

\bibitem[\protect\citeauthoryear{{Orkisz} et~al.,}{{Orkisz}
  et~al.}{2017}]{Orkisz17}
{Orkisz} J.~H.,  et~al., 2017, \mn@doi [\aap] {10.1051/0004-6361/201629220},
  \href {http://adsabs.harvard.edu/abs/2017A%26A...599A..99O} {599, A99}

\bibitem[\protect\citeauthoryear{{Parsons} et~al.,}{{Parsons}
  et~al.}{2018}]{Parsons18}
{Parsons} H.,  et~al., 2018, \mn@doi [\apjs] {10.3847/1538-4365/aa989c}, \href
  {http://adsabs.harvard.edu/abs/2018ApJS..234...22P} {234, 22}

\bibitem[\protect\citeauthoryear{{Pe{\~n}aloza}, {Clark}, {Glover}, {Shetty}
  \& {Klessen}}{{Pe{\~n}aloza} et~al.}{2017}]{Penaloza17}
{Pe{\~n}aloza} C.~H.,  {Clark} P.~C.,  {Glover} S.~C.~O.,  {Shetty} R.,
  {Klessen} R.~S.,  2017, \mn@doi [\mnras] {10.1093/mnras/stw2892}, \href
  {http://adsabs.harvard.edu/abs/2017MNRAS.465.2277P} {465, 2277}

\bibitem[\protect\citeauthoryear{{Pe{\~n}aloza}, {Clark}, {Glover}  \&
  {Klessen}}{{Pe{\~n}aloza} et~al.}{2018}]{Penaloza18}
{Pe{\~n}aloza} C.~H.,  {Clark} P.~C.,  {Glover} S.~C.~O.,   {Klessen} R.~S.,
  2018, \mn@doi [\mnras] {10.1093/mnras/stx3263}, \href
  {http://adsabs.harvard.edu/abs/2018MNRAS.475.1508P} {475, 1508}

\bibitem[\protect\citeauthoryear{{Planck Collaboration} et~al.,}{{Planck
  Collaboration} et~al.}{2014}]{Planck14}
{Planck Collaboration} et~al., 2014, \mn@doi [\aap]
  {10.1051/0004-6361/201323195}, \href
  {http://adsabs.harvard.edu/abs/2014A%26A...571A..11P} {571, A11}

\bibitem[\protect\citeauthoryear{{Ragan}, {Henning}, {Tackenberg}, {Beuther},
  {Johnston}, {Kainulainen}  \& {Linz}}{{Ragan} et~al.}{2014}]{Ragan14}
{Ragan} S.~E.,  {Henning} T.,  {Tackenberg} J.,  {Beuther} H.,  {Johnston}
  K.~G.,  {Kainulainen} J.,   {Linz} H.,  2014, \mn@doi [\aap]
  {10.1051/0004-6361/201423401}, \href
  {http://adsabs.harvard.edu/abs/2014A%26A...568A..73R} {568, A73}

\bibitem[\protect\citeauthoryear{{Ragan}, {Moore}, {Eden}, {Hoare}, {Elia}  \&
  {Molinari}}{{Ragan} et~al.}{2016}]{Ragan16}
{Ragan} S.~E.,  {Moore} T.~J.~T.,  {Eden} D.~J.,  {Hoare} M.~G.,  {Elia} D.,
  {Molinari} S.,  2016, \mn@doi [\mnras] {10.1093/mnras/stw1870}, \href
  {http://adsabs.harvard.edu/abs/2016MNRAS.462.3123R} {462, 3123}

\bibitem[\protect\citeauthoryear{{Ragan}, {Moore}, {Eden}, {Hoare}, {Urquhart},
  {Elia}  \& {Molinari}}{{Ragan} et~al.}{2018}]{Ragan18}
{Ragan} S.~E.,  {Moore} T.~J.~T.,  {Eden} D.~J.,  {Hoare} M.~G.,  {Urquhart}
  J.~S.,  {Elia} D.,   {Molinari} S.,  2018, \mn@doi [\mnras]
  {10.1093/mnras/sty1672}, \href
  {http://adsabs.harvard.edu/abs/2018MNRAS.479.2361R} {479, 2361}

\bibitem[\protect\citeauthoryear{{Reid} \& {Brunthaler}}{{Reid} \&
  {Brunthaler}}{2020}]{Reid20}
{Reid} M.~J.,  {Brunthaler} A.,  2020, arXiv e-prints, \href
  {https://ui.adsabs.harvard.edu/abs/2020arXiv200104386R} {p. arXiv:2001.04386}

\bibitem[\protect\citeauthoryear{{Reid}, {Schneps}, {Moran}, {Gwinn}, {Genzel},
  {Downes}  \& {Roennaeng}}{{Reid} et~al.}{1988}]{Reid88}
{Reid} M.~J.,  {Schneps} M.~H.,  {Moran} J.~M.,  {Gwinn} C.~R.,  {Genzel} R.,
  {Downes} D.,   {Roennaeng} B.,  1988, \mn@doi [\apj] {10.1086/166514}, \href
  {https://ui.adsabs.harvard.edu/abs/1988ApJ...330..809R} {330, 809}

\bibitem[\protect\citeauthoryear{{Reid}, {Dame}, {Menten}  \&
  {Brunthaler}}{{Reid} et~al.}{2016}]{Reid2016}
{Reid} M.~J.,  {Dame} T.~M.,  {Menten} K.~M.,   {Brunthaler} A.,  2016, \mn@doi
  [\apj] {10.3847/0004-637X/823/2/77}, \href
  {http://adsabs.harvard.edu/abs/2016ApJ...823...77R} {823, 77}

\bibitem[\protect\citeauthoryear{{Rigby} et~al.,}{{Rigby}
  et~al.}{2016}]{Rigby16}
{Rigby} A.~J.,  et~al., 2016, \mn@doi [\mnras] {10.1093/mnras/stv2808}, \href
  {http://adsabs.harvard.edu/abs/2016MNRAS.456.2885R} {456, 2885}

\bibitem[\protect\citeauthoryear{{Rigby} et~al.,}{{Rigby}
  et~al.}{2019}]{Rigby19}
{Rigby} A.~J.,  et~al., 2019, \mn@doi [\aap] {10.1051/0004-6361/201935236},
  \href {https://ui.adsabs.harvard.edu/abs/2019A&A...632A..58R} {632, A58}

\bibitem[\protect\citeauthoryear{{Rodriguez-Fernandez}, {Combes},
  {Martin-Pintado}, {Wilson}  \& {Apponi}}{{Rodriguez-Fernandez}
  et~al.}{2006}]{Rodriguez-Fernandez06}
{Rodriguez-Fernandez} N.~J.,  {Combes} F.,  {Martin-Pintado} J.,  {Wilson}
  T.~L.,   {Apponi} A.,  2006, \mn@doi [\aap] {10.1051/0004-6361:20064813},
  \href {http://adsabs.harvard.edu/abs/2006A%26A...455..963R} {455, 963}

\bibitem[\protect\citeauthoryear{{Rosolowsky} et~al.,}{{Rosolowsky}
  et~al.}{2010}]{Rosolowsky10}
{Rosolowsky} E.,  et~al., 2010, \mn@doi [\apjs] {10.1088/0067-0049/188/1/123},
  \href {https://ui.adsabs.harvard.edu/abs/2010ApJS..188..123R} {188, 123}

\bibitem[\protect\citeauthoryear{{Sanna} et~al.,}{{Sanna}
  et~al.}{2014}]{Sanna2014}
{Sanna} A.,  et~al., 2014, \mn@doi [\apj] {10.1088/0004-637X/781/2/108}, \href
  {http://adsabs.harvard.edu/abs/2014ApJ...781..108S} {781, 108}

\bibitem[\protect\citeauthoryear{{Schinnerer} et~al.,}{{Schinnerer}
  et~al.}{2017}]{Schinnerer17}
{Schinnerer} E.,  et~al., 2017, \mn@doi [\apj] {10.3847/1538-4357/836/1/62},
  \href {http://adsabs.harvard.edu/abs/2017ApJ...836...62S} {836, 62}

\bibitem[\protect\citeauthoryear{{Schisano} et~al.,}{{Schisano}
  et~al.}{2014}]{Schisano14}
{Schisano} E.,  et~al., 2014, \mn@doi [\apj] {10.1088/0004-637X/791/1/27},
  \href {http://adsabs.harvard.edu/abs/2014ApJ...791...27S} {791, 27}

\bibitem[\protect\citeauthoryear{{Schruba}, {Leroy}, {Walter}, {Sandstrom}  \&
  {Rosolowsky}}{{Schruba} et~al.}{2010}]{Schruba10}
{Schruba} A.,  {Leroy} A.~K.,  {Walter} F.,  {Sandstrom} K.,   {Rosolowsky} E.,
   2010, \mn@doi [\apj] {10.1088/0004-637X/722/2/1699}, \href
  {https://ui.adsabs.harvard.edu/abs/2010ApJ...722.1699S} {722, 1699}

\bibitem[\protect\citeauthoryear{{Schuller} et~al.,}{{Schuller}
  et~al.}{2017}]{Schuller17}
{Schuller} F.,  et~al., 2017, \mn@doi [\aap] {10.1051/0004-6361/201628933},
  \href {https://ui.adsabs.harvard.edu/abs/2017A&A...601A.124S} {601, A124}

\bibitem[\protect\citeauthoryear{{Sjouwerman}, {Lindqvist}, {van Langevelde}
  \& {Diamond}}{{Sjouwerman} et~al.}{2002}]{Sjouwerman02}
{Sjouwerman} L.~O.,  {Lindqvist} M.,  {van Langevelde} H.~J.,   {Diamond}
  P.~J.,  2002, \mn@doi [\aap] {10.1051/0004-6361:20020983}, \href
  {https://ui.adsabs.harvard.edu/abs/2002A%26A...391..967S} {391, 967}

\bibitem[\protect\citeauthoryear{{Smartt} \& {Rolleston}}{{Smartt} \&
  {Rolleston}}{1997}]{Smartt97}
{Smartt} S.~J.,  {Rolleston} W.~R.~J.,  1997, \mn@doi [\apjl] {10.1086/310640},
  \href {http://adsabs.harvard.edu/abs/1997ApJ...481L..47S} {481, L47}

\bibitem[\protect\citeauthoryear{{Sofue} \& {Nakanishi}}{{Sofue} \&
  {Nakanishi}}{2016}]{Sofue16}
{Sofue} Y.,  {Nakanishi} H.,  2016, \mn@doi [\pasj] {10.1093/pasj/psw062},
  \href {http://adsabs.harvard.edu/abs/2016PASJ...68...63S} {68, 63}

\bibitem[\protect\citeauthoryear{{Sormani} \& {Barnes}}{{Sormani} \&
  {Barnes}}{2019}]{Sormani19}
{Sormani} M.~C.,  {Barnes} A.~T.,  2019, \mn@doi [\mnras]
  {10.1093/mnras/stz046}, \href
  {http://adsabs.harvard.edu/abs/2019MNRAS.484.1213S} {484, 1213}

\bibitem[\protect\citeauthoryear{{Sormani}, {Binney}  \& {Magorrian}}{{Sormani}
  et~al.}{2015a}]{Sormani15}
{Sormani} M.~C.,  {Binney} J.,   {Magorrian} J.,  2015a, \mn@doi [\mnras]
  {10.1093/mnras/stv441}, \href
  {http://adsabs.harvard.edu/abs/2015MNRAS.449.2421S} {449, 2421}

\bibitem[\protect\citeauthoryear{{Sormani}, {Binney}  \& {Magorrian}}{{Sormani}
  et~al.}{2015b}]{Sormani15c}
{Sormani} M.~C.,  {Binney} J.,   {Magorrian} J.,  2015b, \mn@doi [\mnras]
  {10.1093/mnras/stv2067}, \href
  {http://adsabs.harvard.edu/abs/2015MNRAS.454.1818S} {454, 1818}

\bibitem[\protect\citeauthoryear{{Sormani}, {Tre{\ss}}, {Ridley}, {Glover},
  {Klessen}, {Binney}, {Magorrian}  \& {Smith}}{{Sormani}
  et~al.}{2018}]{Sormani18}
{Sormani} M.~C.,  {Tre{\ss}} R.~G.,  {Ridley} M.,  {Glover} S.~C.~O.,
  {Klessen} R.~S.,  {Binney} J.,  {Magorrian} J.,   {Smith} R.,  2018, \mn@doi
  [\mnras] {10.1093/mnras/stx3258}, \href
  {http://adsabs.harvard.edu/abs/2018MNRAS.475.2383S} {475, 2383}

\bibitem[\protect\citeauthoryear{{Sormani} et~al.,}{{Sormani}
  et~al.}{2019}]{Sormani19a}
{Sormani} M.~C.,  et~al., 2019, \mn@doi [\mnras] {10.1093/mnras/stz2054}, \href
  {https://ui.adsabs.harvard.edu/abs/2019MNRAS.488.4663S} {488, 4663}

\bibitem[\protect\citeauthoryear{{Sormani}, {Tress}, {Glover}, {Klessen},
  {Battersby}, {Clark}, {Hatchfield}  \& {Smith}}{{Sormani}
  et~al.}{2020}]{Sormani20}
{Sormani} M.~C.,  {Tress} R.~G.,  {Glover} S. C.~O.,  {Klessen} R.~S.,
  {Battersby} C.~D.,  {Clark} P.~C.,  {Hatchfield} H.~P.,   {Smith} R.~J.,
  2020, \mn@doi [\mnras] {10.1093/mnras/staa1999}, \href
  {https://ui.adsabs.harvard.edu/abs/2020MNRAS.tmp.2115S} {}

\bibitem[\protect\citeauthoryear{{Stark} \& {Bania}}{{Stark} \&
  {Bania}}{1986}]{Stark86}
{Stark} A.~A.,  {Bania} T.~M.,  1986, \mn@doi [\apjl] {10.1086/184695}, \href
  {http://adsabs.harvard.edu/abs/1986ApJ...306L..17S} {306, L17}

\bibitem[\protect\citeauthoryear{{Stark} \& {Lee}}{{Stark} \&
  {Lee}}{2006}]{Stark06}
{Stark} A.~A.,  {Lee} Y.,  2006, \mn@doi [\apjl] {10.1086/504036}, \href
  {http://adsabs.harvard.edu/abs/2006ApJ...641L.113S} {641, L113}

\bibitem[\protect\citeauthoryear{{Su} et~al.,}{{Su} et~al.}{2019}]{Su19}
{Su} Y.,  et~al., 2019, \mn@doi [\apjs] {10.3847/1538-4365/aaf1c8}, \href
  {https://ui.adsabs.harvard.edu/abs/2019ApJS..240....9S} {240, 9}

\bibitem[\protect\citeauthoryear{{Suwannajak}, {Tan}  \& {Leroy}}{{Suwannajak}
  et~al.}{2014}]{Suwannajak14}
{Suwannajak} C.,  {Tan} J.~C.,   {Leroy} A.~K.,  2014, \mn@doi [\apj]
  {10.1088/0004-637X/787/1/68}, \href
  {https://ui.adsabs.harvard.edu/abs/2014ApJ...787...68S} {787, 68}

\bibitem[\protect\citeauthoryear{{Suzuki}, {Fukui}, {Torii}, {Machida}  \&
  {Matsumoto}}{{Suzuki} et~al.}{2015}]{Suzuki10}
{Suzuki} T.~K.,  {Fukui} Y.,  {Torii} K.,  {Machida} M.,   {Matsumoto} R.,
  2015, \mn@doi [\mnras] {10.1093/mnras/stv2188}, \href
  {https://ui.adsabs.harvard.edu/abs/2015MNRAS.454.3049S} {454, 3049}

\bibitem[\protect\citeauthoryear{{Sz{\H u}cs}, {Glover}  \& {Klessen}}{{Sz{\H
  u}cs} et~al.}{2014}]{Szucs14}
{Sz{\H u}cs} L.,  {Glover} S.~C.~O.,   {Klessen} R.~S.,  2014, \mn@doi [\mnras]
  {10.1093/mnras/stu2013}, \href
  {http://adsabs.harvard.edu/abs/2014MNRAS.445.4055S} {445, 4055}

\bibitem[\protect\citeauthoryear{{Tanaka}}{{Tanaka}}{2018}]{Tanaka18}
{Tanaka} K.,  2018, \mn@doi [\apj] {10.3847/1538-4357/aabd77}, \href
  {https://ui.adsabs.harvard.edu/abs/2018ApJ...859...86T} {859, 86}

\bibitem[\protect\citeauthoryear{{Tanaka}, {Kamegai}, {Nagai}  \&
  {Oka}}{{Tanaka} et~al.}{2007}]{Tanaka07}
{Tanaka} K.,  {Kamegai} K.,  {Nagai} M.,   {Oka} T.,  2007, \mn@doi [\pasj]
  {10.1093/pasj/59.2.323}, \href
  {http://adsabs.harvard.edu/abs/2007PASJ...59..323T} {59, 323}

\bibitem[\protect\citeauthoryear{{Thompson}, {Urquhart}, {Moore}  \&
  {Morgan}}{{Thompson} et~al.}{2012}]{Thompson12}
{Thompson} M.~A.,  {Urquhart} J.~S.,  {Moore} T.~J.~T.,   {Morgan} L.~K.,
  2012, \mn@doi [\mnras] {10.1111/j.1365-2966.2011.20315.x}, \href
  {http://adsabs.harvard.edu/abs/2012MNRAS.421..408T} {421, 408}

\bibitem[\protect\citeauthoryear{{Tress}, {Sormani}, {Glover}, {Klessen},
  {Battersby}, {Clark}, {Hatchfield}  \& {Smith}}{{Tress}
  et~al.}{2020}]{Tress20}
{Tress} R.~G.,  {Sormani} M.~C.,  {Glover} S. C.~O.,  {Klessen} R.~S.,
  {Battersby} C.~D.,  {Clark} P.~C.,  {Hatchfield} H.~P.,   {Smith} R.~J.,
  2020, arXiv e-prints, \href
  {https://ui.adsabs.harvard.edu/abs/2020arXiv200406724T} {p. arXiv:2004.06724}

\bibitem[\protect\citeauthoryear{{Tsuboi}, {Kobayashi}, {Ishiguro}  \&
  {Murata}}{{Tsuboi} et~al.}{1991}]{Tsuboi91}
{Tsuboi} M.,  {Kobayashi} H.,  {Ishiguro} M.,   {Murata} Y.,  1991, \pasj,
  \href {https://ui.adsabs.harvard.edu/abs/1991PASJ...43L..27T} {43, L27}

\bibitem[\protect\citeauthoryear{{Umemoto} et~al.,}{{Umemoto}
  et~al.}{2017}]{Umemoto17}
{Umemoto} T.,  et~al., 2017, \mn@doi [\pasj] {10.1093/pasj/psx061}, \href
  {http://adsabs.harvard.edu/abs/2017PASJ...69...78U} {69, 78}

\bibitem[\protect\citeauthoryear{{Urquhart} et~al.,}{{Urquhart}
  et~al.}{2013}]{Urquhart13}
{Urquhart} J.~S.,  et~al., 2013, \mn@doi [\mnras] {10.1093/mnras/stt287}, \href
  {http://adsabs.harvard.edu/abs/2013MNRAS.431.1752U} {431, 1752}

\bibitem[\protect\citeauthoryear{{Urquhart} et~al.,}{{Urquhart}
  et~al.}{2014}]{Urquhart14}
{Urquhart} J.~S.,  et~al., 2014, \mn@doi [\aap] {10.1051/0004-6361/201424126},
  \href {http://adsabs.harvard.edu/abs/2014A%26A...568A..41U} {568, A41}

\bibitem[\protect\citeauthoryear{{Urquhart} et~al.,}{{Urquhart}
  et~al.}{2018}]{Urquhart18}
{Urquhart} J.~S.,  et~al., 2018, \mn@doi [\mnras] {10.1093/mnras/stx2258},
  \href {https://ui.adsabs.harvard.edu/abs/2018MNRAS.473.1059U} {473, 1059}

\bibitem[\protect\citeauthoryear{{Walker} et~al.,}{{Walker}
  et~al.}{2018}]{Walker18}
{Walker} D.~L.,  et~al., 2018, \mn@doi [\mnras] {10.1093/mnras/stx2898}, \href
  {https://ui.adsabs.harvard.edu/abs/2018MNRAS.474.2373W} {474, 2373}

\bibitem[\protect\citeauthoryear{{Walsh}, {Burton}, {Hyland}  \&
  {Robinson}}{{Walsh} et~al.}{1998}]{Walsh98}
{Walsh} A.~J.,  {Burton} M.~G.,  {Hyland} A.~R.,   {Robinson} G.,  1998,
  \mn@doi [\mnras] {10.1046/j.1365-8711.1998.02014.x}, \href
  {https://ui.adsabs.harvard.edu/abs/1998MNRAS.301..640W} {301, 640}

\bibitem[\protect\citeauthoryear{{Wang} et~al.,}{{Wang} et~al.}{2020}]{Wang20}
{Wang} Y.,  et~al., 2020, \mn@doi [\aap] {10.1051/0004-6361/201937095}, \href
  {https://ui.adsabs.harvard.edu/abs/2020A&A...634A..83W} {634, A83}

\bibitem[\protect\citeauthoryear{{Yusef-Zadeh} et~al.,}{{Yusef-Zadeh}
  et~al.}{2009}]{Yusef-Zadeh09}
{Yusef-Zadeh} F.,  et~al., 2009, \mn@doi [\apj] {10.1088/0004-637X/702/1/178},
  \href {https://ui.adsabs.harvard.edu/abs/2009ApJ...702..178Y} {702, 178}

\bibitem[\protect\citeauthoryear{{Zucker}, {Battersby}  \& {Goodman}}{{Zucker}
  et~al.}{2015}]{Zucker15}
{Zucker} C.,  {Battersby} C.,   {Goodman} A.,  2015, \mn@doi [\apj]
  {10.1088/0004-637X/815/1/23}, \href
  {http://adsabs.harvard.edu/abs/2015ApJ...815...23Z} {815, 23}

\makeatother
\end{thebibliography}

\appendix

\section{ORAC-DR parameters}
\label{recipe_param}
The available recipe parameters are described in the \textsc{REDUCE\_SCIENCE\_NARROWLINE} documentation and summarised in the \textit{Classified Recipe Parameters} appendix of Starlink Cookbook 20\footnote{\texttt{http://www.starlink.ac.uk/devdocs/sc20.htx/sc20.html}}.

We first list the parameters that were constant throughout the survey and will be applied to all $^{12}$CO data in the CHIMPS2 survey.
The following parameters controlled the creation of the spectral cubes with \textsc{smurf:makecube} \citep{Chapin13,Jenness13}, and the maximum size of input data before they were processed in chunks.
\begin{verbatim}
CUBE_WCS = GALACTIC
PIXEL_SCALE = 6.0
SPREAD_METHOD = gauss
SPREAD_WIDTH = 9
SPREAD_FWHM_OR_ZERO = 6  
TILE = 0
CUBE_MAXSIZE = 1536

CHUNKSIZE = 12288
\end{verbatim}

The following parameters controlled the creation of the longitude-velocity maps and spectral-channel re-binning for the tiling of a large number of tiles.
\begin{verbatim}
REBIN = 1.0
LV_IMAGE = 1
LV_AXIS = skylat
LV_ESTIMATOR = sum
\end{verbatim}

To guide the automated rejection of spectra affected by artefacts extraneous noise the following parameters were used.
\begin{verbatim}
BASELINE_LINEARITY = 1
BASELINE_LINEARITY_LINEWIDTH = base
BASELINE_REGIONS = -406.8:-272.0,124.0:377.5
BASELINE_LINEARITY_MINRMS = 0.080
HIGHFREQ_INTERFERENCE = 1
HIGHFREQ_RINGING = 0
LOWFREQ_INTERFERENCE = 1
LOWFREQ_INTERFERENCE_THRESH_CLIP = 4.0
\end{verbatim}

These too were constants, except \texttt{BASELINE\_LINEARITY\_LINEWIDTH} was sometimes set to a range to be excluded from the non-linearity tests if there was a single continuous section of emission, otherwise \texttt{BASELINE\_REGIONS} was used inclusively. \texttt{HIGHFREQ\_RINGING} was only enabled (set to 1) when ringing \citep{Jenness15} was present in HARP Receptor H07. \texttt{LOWFREQ\_INTERFERENCE\_THRESH\_CLIP} was set higher -- 6, 8, or 10 -- as needed for $^{12}$CO observations in the CMZ.

The following three parameters controlled how the receptor-to-receptor flat field was to be determined. The responses are normalised to Receptor H05, except in 15 cases in where H05 had failed quality-assurance criteria and H10 was substituted. In three CMZ cases the index method was preferred, using well-determined flat ratios from the same night.  The regions used to derive the flat field were estimated by averaging all the spectra in the first pass of a reduction, then tuning through border velocity channels until there was deemed to be sufficient signal that was not overly concentrated, typically when the mean flux exceeded 0.2\,K.
\begin{verbatim}
FLATFIELD = 1
FLAT_METHOD = sum
FLAT_REGIONS = -87.0:54.0,90.0:190.0
\end{verbatim}

For $^{12}$CO observations in the CMZ, the following parameters related to the baseline fitting were used.
\begin{verbatim}
BASELINE_METHOD = auto
BASELINE_ORDER = 1
FREQUENCY_SMOOTH = 25
BASELINE_NUMBIN = 128
BASELINE_EMISSION_CLIP = 1.0,1.3,1.6,2.0,2.5
\end{verbatim}
In some cases the baseline order was required to be set to 4.
\begin{verbatim}
BASELINE_ORDER = 4
\end{verbatim}

The velocity coverage of the output data products in the CMZ were determined to be $-$407 to 355, and assigned to the \texttt{FINAL\_LOWER\_VELOCITY} and \texttt{FINAL\_UPPER\_VELOCITY} parameters.

The velocity limits containing all identified emission with a margin for error were set by \texttt{MOMENTS\_LOWER\_VELOCITY} and \texttt{MOMENTS\_UPPER\_VELOCITY} to aid in the creation of moments' maps, such the integrated emission.

The final set of parameters were only applicable when there was noticeable contamination from the reference (off-position).
\begin{verbatim}
CLUMP_METHOD = clumpfind
SUBTRACT_REF_EMISSION = 1
REF_EMISSION_MASK_SOURCE = both
REF_EMISSION_COMBINE_REFPOS = 1
REF_EMISSION_BOXSIZE = 19
\end{verbatim}

\clearpage
\onecolumn

\noindent
Author affiliations:\\
$^{1}$Astrophysics Research Institute, Liverpool John Moores University, IC2, Liverpool Science Park, 146 Brownlow Hill, Liverpool, L3 5RF, UK\\
$^{2}$East Asian Observatory, 660 North A'ohoku Place, Hilo, Hawaii 96720, USA\\
$^{3}$RAL Space, STFC Rutherford Appleton Laboratory, Chilton, Didcot, Oxfordshire OX11 0QX, UK\\
$^{4}$School of Physics and Astronomy, Cardiff University, Queen's Buildings, The Parade, Cardiff CF24 3AA, UK\\
$^{5}$Department of Physics, University of Alberta, Edmonton, AB T6G 2R3, Canada\\
$^{6}$Purple Mountain Observatory and Key Laboratory of Radio Astronomy, Chinese Academy of Sciences, Nanjing 210034, People's Republic of China\\
$^{7}$Korea Astronomy and Space Science Institute, 776 Daedeokdae-ro, Yuseong-gu, Daejon 34055, Republic of Korea\\
$^{8}$University of Science and Technology, Korea (UST), 217 Gajeong-ro, Yuseong-gu, Daejeon 34113, Republic of Korea\\
$^{9}$European Southern Observatory, Karl-Schwarzschild-Strasse 2, D-85748 Garching bei M\"{u}nchen, Germany\\
$^{10}$Institute of Astronomy and Department of Physics, National Tsing Hua University, 300, Hsinchu, Taiwan\\
$^{11}$Nobeyama Radio Observatory, National Astronomical Observatory of Japan, National Institutes of Natural Sciences, 462-2, Nobeyama, Minamimaki, Minamimaksu, Nagano 384-1305, Japan\\
$^{12}$Department of Astronomical Science, School of Physical Science, SOKENDAI (The Graduate University for Advanced Studies), 2-21-1, Osawa, Mitaka, Tokyo 181-8588, Japan\\
$^{13}$Centre for Astrophysics and Planetary Science, University of Kent, Canterbury CT2 7NH, UK\\
$^{14}$Department of Physics \& Astronomy, Kwantlen Polytechnic University, 12666 72nd Avenue, Surrey, BC, V3W 2M8, Canada\\
$^{15}$Wartburg College, Waverly, IA 50677, USA\\
$^{16}$Subaru Telescope, National Astronomical Observatory of Japan, National Institutes of Natural Sciences, 650 North A'ohoku Place, Hilo, HI 96720, USA\\
$^{17}$Centre de Recherche en Astrophysique du Qu\'{e}bec, D\'{e}partement de physique, de g\'{e}nie physique et d'optique, Universit\'{e} Laval, QC G1K 7P4, Canada\\
$^{18}$Key Laboratory for Research in Galaxies and Cosmology, Shanghai Astronomical Observatory, Chinese Academy of Sciences, 80 Nandan Road, Shanghai 200030, People’s Republic of China\\
$^{19}$Department of Astrophysics, Vietnam National Space Center (VNSC), Vietnam Academy of Science and Technology (VAST), 18 Hoang Quoc Viet, Ha Noi, Vietnam\\
$^{20}$School of Space Research, Kyung Hee University, 1732 Deogyeong-daero, Giheung-gu, Yongin-si, Gyeonggi-do 17104, Republic of Korea\\
$^{21}$SUPA, School of Physics and Astronomy, University of St Andrews, North Haugh, St Andrews KY16 9SS, UK\\
$^{22}$National Astronomical Observatories, Chinese Academy of Sciences, 20A Datun Road, Chaoyang District, Beijing 100012, China\\
$^{23}$Key Laboratory of FAST, NAOC, Chinese Academy of Science, Beijing 100012, China\\
$^{24}$Max-Planck-Institut f\"{u}r Astronomie, K\"{o}nigstuhl 17, D-69117 Heidelberg, Germany\\
$^{25}$Division of Particle and Astrophysical Science, Graduate School of Science, Nagoya University, Furo-cho, Chikusa-ku, Nagoya, Aichi 464-8602, Japan\\
$^{26}$Department of Earth and Space Science, Graduate School of Science, Osaka University, 1-1, Machikaneyama-cho, Toyonaka, Osaka 560-0043, Japan\\
$^{27}$Astronomical Institute, Graduate School of Science, Tohoku University, Aoba, Sendai, Miyagi 980-8578, Japan\\
$^{28}$Centre for Astrophysics Research, School of Physics Astronomy \& Mathematics, University of Hertfordshire, College Lane, Hatfield, Herts AL10 9AB, UK\\
$^{29}$SUPA, School of Science and Engineering, University of Dundee, Nethergate, Dundee DD1 4HN, U.K.\\
$^{30}$Department of Physics and Astronomy, University of Waterloo, Waterloo, ONN2L 3G1, Canada\\
$^{31}$Department of Astronomy, Xiamen University, Xiamen, Fujian 361005, China\\
$^{32}$Institute of Astronomy and Astrophysics, Academia Sinica. 11F of Astronomy-Mathematics Building, AS/NTU No.1, Section 4, Roosevelt Rd, Taipei 10617, Taiwan\\
$^{33}$Department of Earth Sciences, National Taiwan Normal University, 88 Section 4, Ting-Chou Road, Taipei 11677, Taiwan\\
$^{34}$Department of Earth Science Education, Seoul National University (SNU), 1 Gwanak-ro, Gwanak-gu, Seoul 08826, Republic of Korea\\
$^{35}$Max-Planck-Institut f\"{u}r Extraterrestrische Physik, Giessenbachstrasse 1, D-85741 Garching bei München, Germany\\
$^{36}$Department of Physics and Astronomy, University of Calgary, 2500 University Drive NW, Calgary, Alberta T2N 1N4, Canada\\
$^{37}$Department of Astronomy, Faculty of Mathematics and Natural Sciences, Institut Teknologi Bandung, Jl. Ganesha 10, Bandung 40132 Indonesia\\
$^{38}$Institute of Astronomy, National Central University, Jhongli 32001, Taiwan\\
$^{39}$Physical Research Laboratory, Navrangpura, Ahmedabad, Gujarat 380009, India\\
$^{40}$SOFIA Science Center, Universities Space Research Association, NASA, Ames Research Center, Moffett Field, CA 94035, USA\\
$^{41}$Xinjiang Astronomical Observatory, Chinese Academy of Sciences, 830011 Urumqi, China\\
$^{42}$IAPS-INAF, via Fosso del Cavaliere 100, I-00133 Rome, Italy\\
$^{43}$Department of Physics and Astronomy, The Open University, Walton Hall, Milton Keynes MK7 6AA, UK\\
$^{44}$Max-Planck-Institut f\"{u}r Radioastronomie, Auf dem H\"{u}gel 69, 53121 Bonn, Germany\\
$^{45}$Department of Physics and Astronomy, McMaster University, 1280 Main St. W, Hamilton, ON L8S 4M1, Canada\\
$^{46}$Astrophysics Group, School of Physics, University of Exeter, Stocker Road, Exeter EX4 4QL, UK\\
$^{47}$Department of Astronomy and Space Science, Chungnam National University, 99 Daehak-ro, Yuseong-gu, Daejeon 34134, Republic of
Korea\\
$^{48}$Jodrell Bank Centre for Astrophysics, School of Physics and Astronomy, University of Manchester, Oxford Road, Manchester M13 9PL, UK\\
$^{49}$School of Physics and Astronomy, University of Leeds, Leeds LS2 9JT, UK\\
$^{50}$National Astronomical Observatory of Japan, 2-21-1, Osawa, Mitaka, Tokyo, 181-8588, Japan\\
$^{51}$Department of Physics and Astronomy, Seoul National University, Seoul 151-747, Republic of Korea\\
$^{52}$Graduate School of Pure and Applied Sciences, University of Tsukuba, 1-1-1 Tennodai, Tsukuba, Ibaraki 305-8571, Japan\\
$^{53}$Tomonaga Center for the History of the Universe, University of Tsukuba, 1-1-1 Tennodai, Tsukuba, Ibaraki 305-8571, Japan\\
$^{54}$Department of Physics and Atmospheric Science, Dalhousie University, Halifax, NS B3H 4R2, Canada\\
$^{55}$Department of Physics, Institute of Science and Technology, Keio University, 3-14-1 Hiyoshi, Kohoku-ku, Yokohama, Kanagawa 223-8522, Japan\\
$^{56}$School of Fundamental Science and Technology, Graduate School of Science and Technology, Keio University, 3-14-1 Hiyoshi, Kohoku-ku, Yokohama, Kanagawa 223-8522, Japan\\
$^{57}$College of Material Science and Chemical Engineering, Hainan University, Hainan 570228, China\\
$^{58}$Key Laboratory of Modern Astronomy and Astrophysics, Nanjing University, Ministry of Education, Nanjing 210093, China\\
$^{59}$Instituto de Radioastronom\'{i}a y Astrof\'{i}sica, Universidad Nacional Auton\'{o}ma de M\'{e}xico, Antigua Carretera a P\'{a}tzcuaro \# 8701 Ex-Hda., San Jos\'{e} de la Huerta, Morelia, Michoac\'{a}n, C.P. 58089, M\'{e}xico\\
$^{60}$Radio Telescope Data Center, Center for Astrophysics, Harvard \& Smithsonian, 60 Garden Street, Cambridge, MA 02138, USA\\
$^{61}$Department of Astronomy, Peking University, 100871 Beijing, China\\
$^{62}$Department of Astronomy, Yunnan University, Key Laboratory of Astroparticle Physics of Yunnan Province, Kunming 650091, China

\bsp
\label{lastpage}

\end{document}